\documentclass[useAMS, usenatbib]{mn2e}

\input psfig.sty

\newcommand {\apj} {ApJ}
\newcommand {\apjl} {ApJL}
\newcommand {\apjs} {ApJS}
\newcommand {\mnras} {MNRAS}

\newcommand {\aap} {A\&A}
\newcommand {\aj} {AJ}

\newcommand {\nat} {Nature}

\newcommand {\etal} {et~al.~}
\def \spose#1{\hbox  to 0pt{#1\hss}}  
\newcommand {\lta} {\mathrel{\spose{\lower 3pt\hbox{$\sim$}}\raise  2.0pt\hbox{$<$}}}
\newcommand {\gta} {\mathrel{\spose{\lower  3pt\hbox{$\sim$}}\raise 2.0pt\hbox{$>$}}}

\newcommand {\ha}  {\ifmmode H\alpha \else H$\alpha $ \fi}

\newcommand {\kms} {\ifmmode  \,\rm km\,s^{-1} \else $\,\rm km\,s^{-1}  $ \fi }
\newcommand {\kpc} {\ifmmode  {\rm kpc}  \else ${\rm  kpc}$ \fi  }  
\newcommand {\pc} {\ifmmode  {\rm pc}  \else ${\rm pc}$ \fi  }  
\newcommand {\Msun} {\ifmmode M_{\odot} \else $M_{\odot}$ \fi} 
\newcommand {\yr} {\ifmmode yr^{-1} \else $yr^{-1}$ \fi} 
\newcommand {\hMsun} {\ifmmode h^{-1}\,\rm M_{\odot} \else $h^{-1}\,\rm M_{\odot}$ \fi}

\newcommand {\LCDM} {\ifmmode \Lambda{\rm CDM} \else $\Lambda{\rm CDM}$ \fi}

\newcommand {\Rvir} {\ifmmode R_{\rm vir} \else $R_{\rm vir}$ \fi}
\newcommand {\Vvir} {\ifmmode V_{\rm  vir} \else  $V_{\rm vir}$  \fi} 
\newcommand {\Mvir} {\ifmmode M_{\rm  vir} \else $M_{\rm  vir}$ \fi}  
\newcommand {\Jvir} {\ifmmode J_{\rm vir} \else $J_{\rm vir}$ \fi} 
\newcommand {\Evir} {\ifmmode E_{\rm vir} \else $E_{\rm vir}$ \fi} 
\newcommand {\lamgal} {\ifmmode \lambda_{\rm gal}  \else $\lambda_{\rm gal}$ \fi} 
\newcommand {\lamstar} {\ifmmode \lambda_{\rm star}  \else $\lambda_{\rm star}$ \fi} 
\newcommand {\lam} {\ifmmode \lambda  \else $\lambda$ \fi} 

\newcommand {\mgal}    {\ifmmode m_{\rm gal}    \else $m_{\rm gal}$ \fi} 
\newcommand {\Mstar}    {\ifmmode M_{\rm star}    \else $M_{\rm star}$ \fi} 
\newcommand {\Mgas}    {\ifmmode M_{\rm gas}    \else $M_{\rm gas}$ \fi} 
\newcommand {\Mgal}  {\ifmmode M_{\rm gal}  \else $M_{\rm gal}$ \fi}
\newcommand {\Matom}  {\ifmmode M_{\rm atom}  \else $M_{\rm atom}$ \fi}
\newcommand {\Mmole}  {\ifmmode M_{\rm mol}  \else $M_{\rm mol}$ \fi}
\newcommand {\jgal} {\ifmmode j_{\rm gal} \else $j_{\rm gal}$ \fi}  
\newcommand {\Jgal} {\ifmmode J_{\rm gal} \else $J_{\rm gal}$ \fi}  
\newcommand {\Vrot} {\ifmmode  V_{\rm rot} \else $V_{\rm rot}$ \fi} 
\newcommand {\Vmax} {\ifmmode  V_{\rm max} \else $V_{\rm max}$ \fi} 

\newcommand {\mdotcool} {\ifmmode  \dot{M}_{\rm cool} \else $\dot{M}_{\rm cool}$ \fi} 
\newcommand {\mdotbar} {\ifmmode  \dot{M}_{\rm bar} \else $\dot{M}_{\rm bar}$ \fi}

\title[Evolution of velocity-mass-size relations] {On the
  Evolution of the Velocity-Mass-Size Relations of Disk-Dominated Galaxies
  over the Past 10 Billion Years.}

\author[Dutton \etal]  {Aaron  A.  Dutton$^{1,2}$\thanks{dutton@uvic.ca}\thanks{CITA National Fellow}, 
  Frank C. van den Bosch$^{3,4}$, 
  Sandra M. Faber$^2$, Luc Simard$^5$, 
 \newauthor{Susan A. Kassin$^6$, David C. Koo$^2$, 
  Kevin Bundy$^7$\thanks{Hubble Fellow}, Jiasheng Huang$^8$, }
 \newauthor{Benjamin J. Weiner$^9$, Michael C.\ Cooper$^9$\thanks{Spitzer Fellow},
   Jeffrey A. Newman$^{10}$, 
Mark Mozena$^2$,}
 \newauthor{Anton M. Koekemoer$^{11}$.}\\
  $^1$Deptartment of Physics and Astronomy, University of Victoria, Victoria, 
      BC, V8P 5C2, Canada.\\
  $^2$UCO/Lick Observatory and Department of Astronomy \& Astrophysics, 
      University of California, Santa Cruz, CA 95064, USA.\\
  $^3$Department of Physics and Astronomy, University of Utah, 115 South 
      1400 East, Salt Lake City, UT 84112-0830, USA.\\
  $^4$Astronomy Department, Yale University, P.O. Box 208101, New Haven, CT 06520-8101, USA.\\ 
  $^5$Herzberg Institute of Astrophysics, National Research Council of
      Canada, 5071 West Saanich Road, Victoria, BC, V9E 2E7, Canada.\\
  $^6$Department of Astrophysics, University of Oxford, Oxford OX1 3RH, UK.\\
  $^7$Astronomy Department, University of California, Berkeley, CA 94705, USA.\\
  $^8$Harvard-Smithsonian Center for Astrophysics, 60 Garden Street, Cambridge, MA 02138, USA.\\
  $^9$Steward Observatory, University of Arizona, 933 N.\ Cherry Avenue, 
      Tucson, AZ 85721 USA.\\
  $^{10}$Department of Physics and Astronomy, University of Pittsburgh, 
      401-C Allen Hall, 3941 O'Hara Street, Pittsburgh, PA 15260, USA.\\
  $^{11}$Space Telescope Science Institute, 3700 San Martin Drive, Baltimore, MD 21218, USA.\\
\vspace{-1.0cm}
}

\begin{document}
             
\date{Accepted 2010 August 17}
             
\pagerange{\pageref{firstpage}--\pageref{lastpage}}\pubyear{2010}

\maketitle           

\label{firstpage}


\begin{abstract}
  We study the evolution of the scaling relations between maximum
  circular velocity, stellar mass and optical half-light radius of
  star-forming disk-dominated galaxies in the context of
  $\Lambda$CDM-based galaxy formation models.  Using data from the
  literature combined with new data from the DEEP2 and AEGIS surveys
  we show that there is a consistent picture for the evolution of
  these scaling relations from $z\sim 2$ to $z=0$, both
  observationally and theoretically. The evolution of the observed
  stellar scaling relations is weaker than that of the virial scaling
  relations of dark matter haloes, which can be reproduced, both
  qualitatively and quantitatively, with a simple,
  cosmologically-motivated model for disk evolution inside growing
  Navarro-Frenk-White dark matter haloes.  In this model optical
  half-light radii are smaller, both at fixed stellar mass and maximum
  circular velocity, at higher redshifts.  This model also predicts
  that the scaling relations between baryonic quantities (baryonic
  mass, baryonic half-mass radii, and maximum circular velocity)
  evolve even more weakly than the corresponding stellar relations.
  We emphasize, though, that this weak evolution does not imply that
  individual galaxies evolve weakly. On the contrary, individual
  galaxies grow strongly in mass, size and velocity, but in such a way
  that they move largely along the scaling relations. Finally, recent
  observations have claimed surprisingly large sizes for a number of
  star-forming disk galaxies at $z \simeq 2$, which has caused some
  authors to suggest that high redshift disk galaxies have abnormally
  high spin parameters. However, we argue that the disk scale lengths
  in question have been systematically overestimated by a factor $\sim
  2$, and that there is an offset of a factor $\sim 1.4$ between
  H$\alpha$ sizes and optical sizes. Taking these effects into
  account, there is no indication that star forming galaxies at high
  redshifts ($z\simeq 2$) have abnormally high spin parameters.
\end{abstract}

\begin{keywords}
  galaxies: evolution -- galaxies: formation -- galaxies: fundamental
  parameters -- galaxies: haloes -- galaxies: high redshift -- galaxies: spiral 
\end{keywords}

\setcounter{footnote}{1}


\section{Introduction}
\label{sec:intro}

In the current paradigm of galaxy formation, galaxy disks are
considered to form from the accretion of gas inside hierarchically
growing cold dark matter (CDM) haloes (White \& Rees 1978). The dark
matter (DM) and gas acquire angular momentum via tidal torques in the
early universe (Peebles 1969). When the gas accretes onto the central
galaxy, this angular momentum may eventually halt the collapse and
lead to the formation of a rotationally supported disk (Fall \&
Efstathiou 1980). Under the assumption that the total specific angular
momentum of the precollapse gas is similar to that of the DM and is
conserved during collapse, this picture leads to predictions of
present-day disk sizes that are in reasonable agreement with
observations (Blumenthal \etal 1984; Dalcanton \etal 1997; Mo, Mao, \&
White 1998; Firmani \& Avila-Reese 2000; de Jong \& Lacey 2000;
Pizagno \etal 2005; Dutton \etal 2007).

The observational relation between the sizes of galaxy disks and their
characteristic rotation velocities (the $RV$ relation) provides a
measure of the specific angular momentum of galaxy disks (e.g.,
Navarro \& Steinmetz 2000).  In the simplest model for disk galaxy
evolution, the sizes and rotation velocities of galaxy disks are
proportional to the sizes and circular velocities of their host dark
matter haloes, and the evolution of disk sizes at fixed circular
velocity scales inversely with the Hubble parameter (e.g., Mo, Mao, \&
White 1998).  For a $\Lambda$CDM cosmology (with $\Omega_{\rm M}=0.3,
\Omega_{\Lambda}=0.7$) this simple model predicts that disks should be
a factor of $\simeq 1.8$ smaller, at fixed circular velocity, at
$z=1$, compared to $z=0$, and a factor of $\simeq 3.0$ smaller at
$z=2$. For a standard CDM cosmology (with $\Omega_{\rm M}=1,
\Omega_{\Lambda}=0$) the predicted evolution is even stronger: a
factor of $\simeq 2.8$ to $z=1$ and a factor of $\simeq 5.2$ to $z=2$.

Early observations indicated evolution in the $RV$ relation with a
scaling of $\sim (1+z)^{-1}$ out to $z\sim 1$, in agreement with
theoretical expectations (Mao, Mo \& White 1998). More recently,
Bouch{\'e} \etal (2007) found that the disk scale size-rotation
velocity relation at $z\simeq 2.2$ has roughly the same zero point as
at $z=0$, in disagreement with simple theoretical expectations.
While there are concerns about sample selection, small statistics,
errors at high-$z$, etc., it is nonetheless appropriate to start
thinking about the implications of this result for galaxy formation.
A proposed explanation for this non-evolution is that disk galaxies at
$z\simeq 2.2$ have spin parameters\footnote{The spin parameter is a
  dimensionless measure of the specific angular momentum of dark
  matter halos.}  a factor of 3 higher than those predicted by
$\Lambda$CDM (Bouch{\'e} \etal 2007; Burkert \etal 2009).  If
confirmed, this could imply a significant modification of the current
paradigm of galaxy disk formation, either through a decoupling of the
angular momentum of baryons and dark matter, or more drastically by
changing the mechanism by which galactic angular momentum is
generated.

A decoupling between the angular momentum of baryons and dark matter
haloes is actually seen in cosmological hydrodynamical simulations of
disk galaxy formation (e.g., Navarro \& Steinmetz 2000; Sales \etal
2009; Piontek \& Steinmetz 2009). However, the result is that baryons
lose angular momentum to the halo, which only makes high disk spin
parameters harder to explain. A possible mechanism for increasing the
specific angular momentum of disk galaxies is the removal of material
with low specific angular momentum though galactic outflows. This
mechanism has been shown theoretically to be effective in dwarf
galaxies (Maller \& Dekel 2002; Dutton 2009; Governato \etal
2010). Galactic outflows appear ubiquitous in star forming galaxies at
redshift $z\sim 3$ (Shapley \etal 2003) and $z=1.4$ (Weiner \etal
2009). But whether they are powerful enough to increase the specific
angular momentum of massive ($V_{\rm rot}=200 \kms$) galaxy disks by a
whole factor of 3 remains to be seen.

The relation between galaxy size and stellar mass (the $RM$ relation)
of disk-dominated galaxies also shows only weak evolution to $z=1$
(Barden \etal 2005). This weak evolution can be understood in the
\LCDM scenario as a consequence of the increase of dark halo
concentrations since $z=1$ (Somerville \etal 2008) and also because
individual disk galaxies tend to evolve roughly {\it along} the $RM$
relation (Firmani \& Avila-Reese 2009).  At higher redshifts, there is
evidence of a factor of $\simeq 2$ decrease in half-light sizes of
disk-dominated galaxies at fixed stellar mass between $z=0$ and
$z\simeq 2.5$ (Trujillo \etal 2006). Again, this evolution is weaker
than the evolution in the halo $RM$ relation, but it is in agreement
with the models of Somerville \etal (2008) and Firmani \& Avila-Reese
(2009). We note that elliptical galaxies have been found to evolve
even more strongly with redshift (e.g., Trujillo \etal 2006; van Dokkum
\etal 2008), and thus there is clear evidence that galaxies of all
types were smaller, at fixed stellar mass, at higher redshifts.

It is possible that the non-evolution of the $RV$ relation and
evolution of the $RM$ relation of disk-galaxies are consistent. This
would require evolution in the $MV$ relation (i.e., the stellar mass
Tully-Fisher relation, Tully \& Fisher 1977), with higher $M$, at
fixed $V$, at higher $z$. However, the opposite has been claimed, with
$\simeq 0.4$ dex lower $M_{\rm star}$ at fixed $V_{\rm max}$ at
$z\simeq 2.2$ compared to $z=0$ (Cresci \etal 2009).  Thus, the
various data sets for the evolution of the $VMR$ relations are
inconsistent.

In this paper, we examine the $VMR$ relations of disk galaxies using
data from the literature as well as new results from the Deep
Extragalactic Evolutionary Probe 2 survey (DEEP2, Davis \etal 2003;
Newman \etal in prep).  We discuss, and resolve, sources of
discrepancies among the various data sets, and compare the data to
predictions of a simple $\Lambda$CDM-based disk-galaxy evolution
model. We show that the observations can be reproduced by a model with
constant spin parameter in agreement with predictions from
$\Lambda$CDM, and a constant galaxy mass fraction. Hence there is no
need to invoke abnormally high spin parameters in order to explain the
scaling relations of disk galaxies at $z\sim 2$.  Throughout we assume
a flat $\Lambda$CDM cosmology with $(\Omega_{\rm
  M},\Omega_{\Lambda},h)=(0.3,0.7,0.7)$.


\section{Observations}
\label{sec:obs}
\subsection{Evolution of the stellar mass Tully-Fisher relation}
The relation between stellar mass\footnote{Note that stellar mass
  throughout is total stellar mass, not disk stellar mass.} (or
luminosity) and rotation velocity (or linewidth) is also commonly
known as the Tully-Fisher relation (TF; Tully \& Fisher 1977).  We
define the velocity as the maximum rotation velocity, $V_{\rm max}$.
Using slit spectroscopy from the DEEP1 survey (Vogt \etal 2005; Weiner
\etal 2005), together with optical and $K$-band imaging, Conselice
\etal (2005) found that, at fixed $V_{\rm max}$, stellar masses are
lower by $0.07\pm0.12$ dex at $z\sim 0.45$, and lower by $0.11\pm0.13$
dex at $z\sim 0.85$, compared to the $z=0$ relation from Bell \& de
Jong (2001). This weak evolution from $z\sim 1$ to $z\sim 0$ was
confirmed by Kassin \etal (2007), who studied the stellar mass TF
relation for a larger sample of galaxies from the DEEP2 survey, but
adopting a different velocity indicator, $S_{0.5}=( 0.5 V^2_{\rm rot}
+ \sigma^2)^{1/2}$, which combines ordered motions (i.e., rotation,
$V_{\rm rot}$) and disordered motions (i.e., dispersion, $\sigma$)
(Weiner \etal 2006a). This new parameter allowed inclusion of mergers
and incompletely settled galaxies, which is useful at high redshifts
were galaxies maybe dynamically ``young''.

Here we use data from Kassin \etal (2007) and calculate the evolution
of stellar mass at fixed $S_{0.5}$ for galaxies with $S_{0.5} >
90\kms$ (i.e., $\log_{10} V_{\rm max}/[\kms] > 2.1$) and inclinations
between $45$ and $70$ degrees. As a comparison relation at redshift
$z=0$ we use the TF relation from Bell \& de Jong (2001), corrected to
a Chabrier (2003) IMF as given below in \S\ref{sec:rv}, and with
$V_{\rm rot}$ converted into $S_{0.5}$ assuming $\sigma=0$. We note
that since the galaxies at $z=0$ are rotation dominated, adopting a
more realistic value of $\sigma\sim 10 \kms$ (for the cold atomic
hydrogen disk) for these galaxies will not change the $z=0$ relation
by any significant amount. The median offsets in stellar mass of the
DEEP2 data with respect to the $z=0$ relation, in three redshift bins
from $z=0.2$ to $z=1.1$, are given in Table~\ref{tab:vmz}. These
results are consistent with no evolution or at most a weak decrease in
stellar masses at fixed circular velocity at higher redshifts.

A relatively strong evolution in the stellar mass TF relation since
$z\sim 0.6$ has been reported by Puech \etal (2010). These authors
find an increase of $0.34$ dex in stellar mass at fixed velocity since
$z\sim 0.6$.  However, the same authors find no evidence for evolution
in baryonic mass at fixed velocity (the baryonic Tully-Fisher
relation) over the same redshift range. The large difference between
the evolution of the stellar and baryonic TF relations implies that
gas fractions evolve significantly. However, the inferred gas
fractions of their galaxies at $z\sim 0.6$ are on average $30\%$,
which is not much higher than that of local galaxies of the same
mass. This implies that the evolution of the stellar mass TF relation
should be only of order 0.1 dex different than that of the baryonic TF
relation. The difference in evolution of 0.4 dex found by Puech \etal
(2010) can thus be traced to the use of inconsistent local baryonic
and stellar mass TF relations.

\begin{table}
 \centering
 \caption{Evolution of the stellar mass-velocity relation of star
   forming galaxies relative to $z=0.0$, using DEEP2 data from Kassin
   \etal (2007).}
  \begin{tabular}{ccccccc}
\hline
\hline  
 redshift range & median $z$ & $\Delta \log_{10} M_{\rm star} | S_{0.5}$ & N\\
\hline
 0.2-0.5  & 0.35 & $-0.06\pm0.09 $ & 19\\
 0.5-0.8  & 0.73 & $-0.03\pm0.11 $ & 25\\
 0.8-1.1  & 0.95 & $-0.15\pm0.09 $ & 29\\
\hline
\hline
\label{tab:vmz}
\end{tabular}
\end{table}

At even higher redshifts evidence has been reported of significant
evolution compared to redshift $z=0$.  Using data from SINS
(Spectroscopic Imaging survey in the NIR with SINFONI, F\"orster
Schreiber \etal 2009), Cresci \etal (2009) found that at fixed $V_{\rm
  max}$, stellar masses at redshift $z\simeq2.2$ are lower by
$0.41\pm0.11$ dex compared to $z=0$. Taken together, the observations
indicate that there is a modest evolution in the zero point of the TF
relation out to $z\sim 1$ and a stronger evolution from $z\sim 1$ to
$z\sim 2$.

\subsection{Evolution in the size-stellar mass relation}
\label{sec:rm}
Barden \etal (2005) found that there was little or no evolution in the
circularized\footnote{For sizes measured in elliptical apertures, the
  quoted size, is conventionally the major axis size, $R$. The
  circularized size is then often defined as $\sqrt{b/a}R$, where
  $b/a$ is the minor to major axis ratio.}  optical half-light
size-stellar mass relation of disk dominated galaxies (defined as
having S\'ersic index $n < 2.5$) from redshift $z\simeq 1$ to $z\simeq
0.1$. Trujillo \etal (2006) measured the evolution of the circularized
rest-frame $V$-band half-light radius-stellar mass relation since
$z\simeq 2.5$, finding that disk-galaxies (S\'ersic index $n< 2.5$)
are a factor of $\simeq 2$ smaller, at fixed $\Mstar$, at $z\simeq
2.5$ than at $z=0.1$. Williams \etal (2010) measure the evolution of
the circularized rest-frame $I$-band half-light radii of galaxies with
stellar masses greater than $6.3\times 10^{10}\Msun$ from $z=0.5$ to
$z=2$, finding strong evolution for both star-forming and non
star-forming galaxies. In order to compare to $z=0.1$ we use a mean
$I$-band half-light size of 5.0 kpc. This has been determined using
our SDSS measurements (see \S\ref{sec:sdss}), and applying a
correction of -0.06 dex to go to rest-frame $I$-band (see
\S\ref{sec:ha}).
 
Here we present new results for the evolution of the disk size-stellar
mass relation for blue-cloud disk dominated galaxies from redshifts
$z=1.2$ to $z=0.1$ using high redshift data from DEEP2 (Davis \etal
2003; Newman \etal, in prep) and the All-wavelength Extended Groth
Strip International Survey (AEGIS, Davis \etal 2007), and a low
redshift comparison sample from the Sloan Digital Sky Survey (SDSS,
York \etal 2000). The main difference of our study with respect to
previous studies (e.g. Barden \etal 2005; Trujillo \etal 2006) is our
use of disk sizes, rather than total sizes. We use disk sizes because
our main interest in this paper is the evolution of galaxy
disks. Total sizes, even for disk-dominated galaxies, depend on the
bulge fraction and bulge size, and thus give a measurement that is
more difficult to interpret.

\subsubsection{DEEP2 data}
Disk sizes have been measured using 2D bulge+disk fits using \textsc{gim2d}
(Simard \etal 2002). The bulge component was assumed to have a
deVaucoulers profile (S\'ersic $n=4$), while the disk was assumed to
be exponential (i.e., S\'ersic $n=1$).  The fits were performed
simultaneously on F606W ($V$) and F814W ($I$) single-orbit Hubble
Space Telescope ({\it HST}) Advanced Camera for Surveys (ACS)
images. The bulge and disk sizes are constrained to be the same in
each filter, but the bulge and disk fluxes are free to vary.  This
choice has been made to maximize signal to noise, at the expense of
using bluer rest-frame wavelengths at higher redshifts: The central
rest-frame wavelength of the images (i.e., the average of the F606W
and F814W images) varies from $\simeq500$ nm at $z=0.4$ to $\simeq
320$ nm at $z=1.2$. This may cause a small ($\simeq 10\%$) systematic
overestimation of the sizes at higher redshifts, because disks tend to
have color gradients (see \S\ref{sec:recon}).  The sizes we present in
this paper are major-axis (i.e., elliptical aperture) disk half-light
sizes, i.e., 1.678 times the disk exponential scale length.

We use both spectroscopic and photometric redshifts between $0.2 < z <
1.4$. We use high quality (``z-quality'' $\ge$ 3) spectroscopic
redshifts from DEEP2 (Davis \etal 2003; Newman \etal, in prep), and
photometric redshifts based on optical to Spitzer/IRAC photometry
(J. Huang \etal, in preparation). For the spectro-$z$ sample, stellar
masses were obtained from optical to NIR spectral energy distribution
(SED) fits from Bundy \etal (2006) which assumed a Chabrier (2003)
initial mass function (IMF). For the photo-$z$ sample we use stellar
masses calculated using a relation between $B$-band stellar
mass-to-light ratio and rest-frame $(U-B)$ color, which we calibrate
against the Bundy \etal (2006) masses. Our calibration is similar to
the ones reported in Lin \etal (2007) and Weiner \etal (2009), but
make use of only rest-frame $(U-B)$ colors, rather than $(U-B)$ and
$(B-V)$. We compute rest-frame $(U-B)$ colors and $B$-band
luminosities by applying kcorrect v4.1.4 (Blanton \& Roweis 2007) to
DEEP2 $B,R,I$ magnitudes.

We select blue-cloud galaxies for this study by using the ``green
valley'' in the $(U-B)$ color-stellar mass plane as the division:
\begin{equation}
(U-B) < 1.05 -0.1 ( \log_{10} M_{\rm star}-10).
\end{equation}
In addition we select only disk-dominated galaxies with disk light
fractions greater than 50\%. This additional cut only removes a small
fraction of galaxies because the vast majority of blue galaxies
already have disk fractions greater than 50\%. In order to minimize
the effects of extinction on disk sizes, colors and stellar masses,
while at the same time keeping a significant sample size, we limit our
sample to galaxies with disk minor-to-major axis ratios greater than
0.5 (corresponding to disk inclinations less than 60 degrees, for a
zero thickness disk).

\subsubsection{Magnitude and surface brightness selection effects}

The DEEP2 spectroscopic survey is limited by apparent $R$-band
magnitude of 24.1 AB. This results in a bias against redder galaxies
at low stellar masses (Willmer \etal 2006). However, the stellar
masses we use from Bundy \etal (2006) are limited by apparent $K$-band
magnitude of about 22.5 AB.  This additional $K$-band selection is
close to a selection on stellar mass and thus removes much of the bias
by removing lower-mass blue galaxies from the sample.  In order to
remove any remaining color biases we apply an additional redshift
dependent lower stellar mass limit of $\log_{10} M_{\rm min}=8.5 + z$.
Our final DEEP2 sample consists of $\sim 800$ galaxies with
spectro-$z$'s and $\sim 2000$ galaxies with photo-$z$'s.

A potentially important selection bias for galaxy size evolution
studies is incompleteness at low surface brightness levels, especially
at higher redshifts due to the cosmological surface brightness
dimming, which goes as $(1+z)^4$.  To determine the impact of surface
brightness selection effects it is customary to create model galaxies,
insert them into real data, and then run the same source detection
algorithms that were applied to the real data. For the single {\it}
orbit {\it HST}/ACS F850LP observations from GEMS used by Barden \etal
(2005) and Trujillo \etal (2006), these tests predict that surface
brightness limits are significant for galaxies brighter than the
magnitude limit of the COMBO-17 photometric redshift survey. However,
as shown by Melbourne \etal (2007), typical large spiral galaxies at
$z\sim 1$ have significant amounts of high-surface-brightness
substructure. This makes them much easier to detect and measure
spectroscopic redshifts for than smooth disks.  In this paper we use
galaxies detected in an effective exposure of two-orbits of {\it
  HST}/ACS imaging (single-orbit each of F606W and F814W). Based on
the results of Melbourne \etal (2007) we conclude that, at the depth
of our images, we are not missing significant numbers of large star
forming galaxies with stellar masses greater than $\sim 10^{10}\Msun$
at $z\sim 1$.

At the other extreme, small galaxies can be confused with stars,
especially with ground based seeing limited imaging. However, stars
can be reliably separated from galaxies based on the fact that stars
tend to be brighter than compact high-redshift galaxies, and also
occupy a different locus from galaxies in ($B-R$, $R-I$) space (Newman
\etal in prep). Furthermore, there is nothing to prevent small
galaxies from being in our photo-$z$ sample. The good agreement
between the size-mass relations from our spectro-$z$ and photo-$z$
samples vindicates the DEEP2 star-galaxy separation method.  We thus
conclude that our sample of star-forming galaxies from DEEP2 is
unlikely to be affected by biases against very small or very large
galaxies above the (redshift dependent) stellar mass limits of our
sample.

\begin{figure*}
  \centerline{ 
\psfig{figure=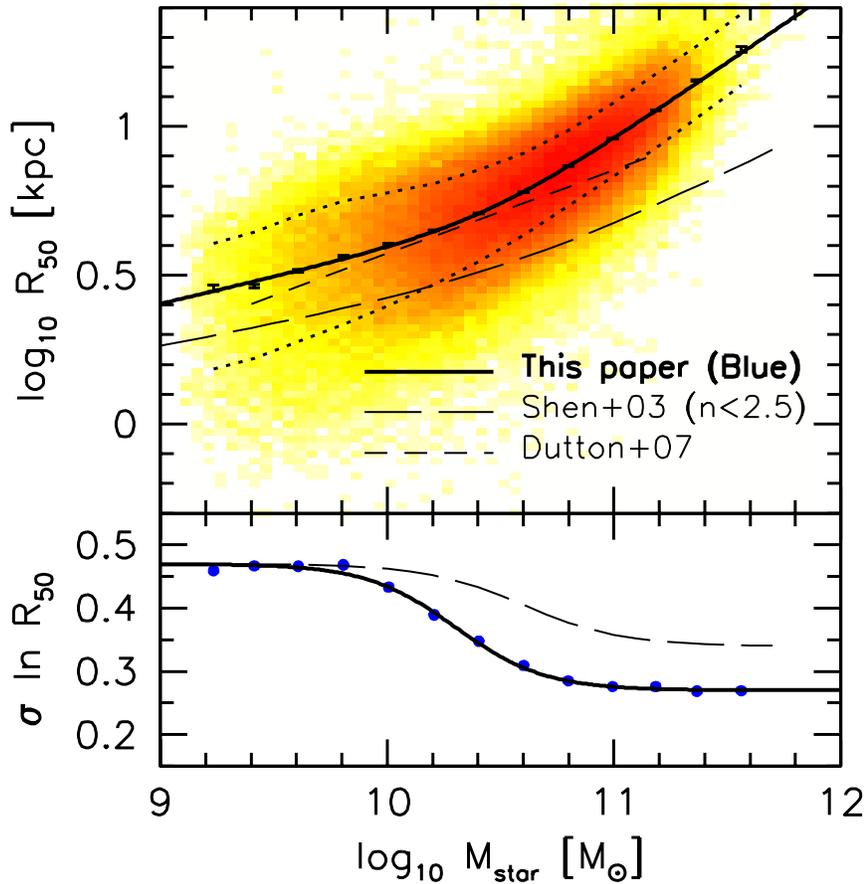,width=0.65\textwidth}}
\caption{ Disk half-light radius-stellar mass ($RM$) relation of
  blue-cloud disk-dominated galaxies (bulge fraction less than 0.5)
  from the SDSS.  The color corresponds to the observed number density
  of galaxies on a logarithmic scale. The error bars show the median
  and error on the median in stellar mass bins of width 0.2 dex. The
  dotted lines show the 84th and 16th percentiles of the size
  distribution. The solid line shows our fit to the median relation
  using Eq.~\ref{eq:rm}. For comparison we also show the total
  half-light radius-stellar mass relation of disk-dominated (S\'ersic
  $n<2.5$) galaxies from Shen \etal (2003, long-dashed line) and the
  disk half-light radius-stellar mass relation from Dutton \etal
  (2007, short-dashed line). The scatter in the size-mass relation is
  given in the lower panel. The points show our measurements from
  SDSS, the solid line shows our fit to these data using
  Eq.~\ref{eq:srm}, while the long-dashed line shows the result from
  Shen \etal (2003).}
\label{fig:rmsdss}
\end{figure*}

\begin{figure*}
\centerline{
\psfig{figure=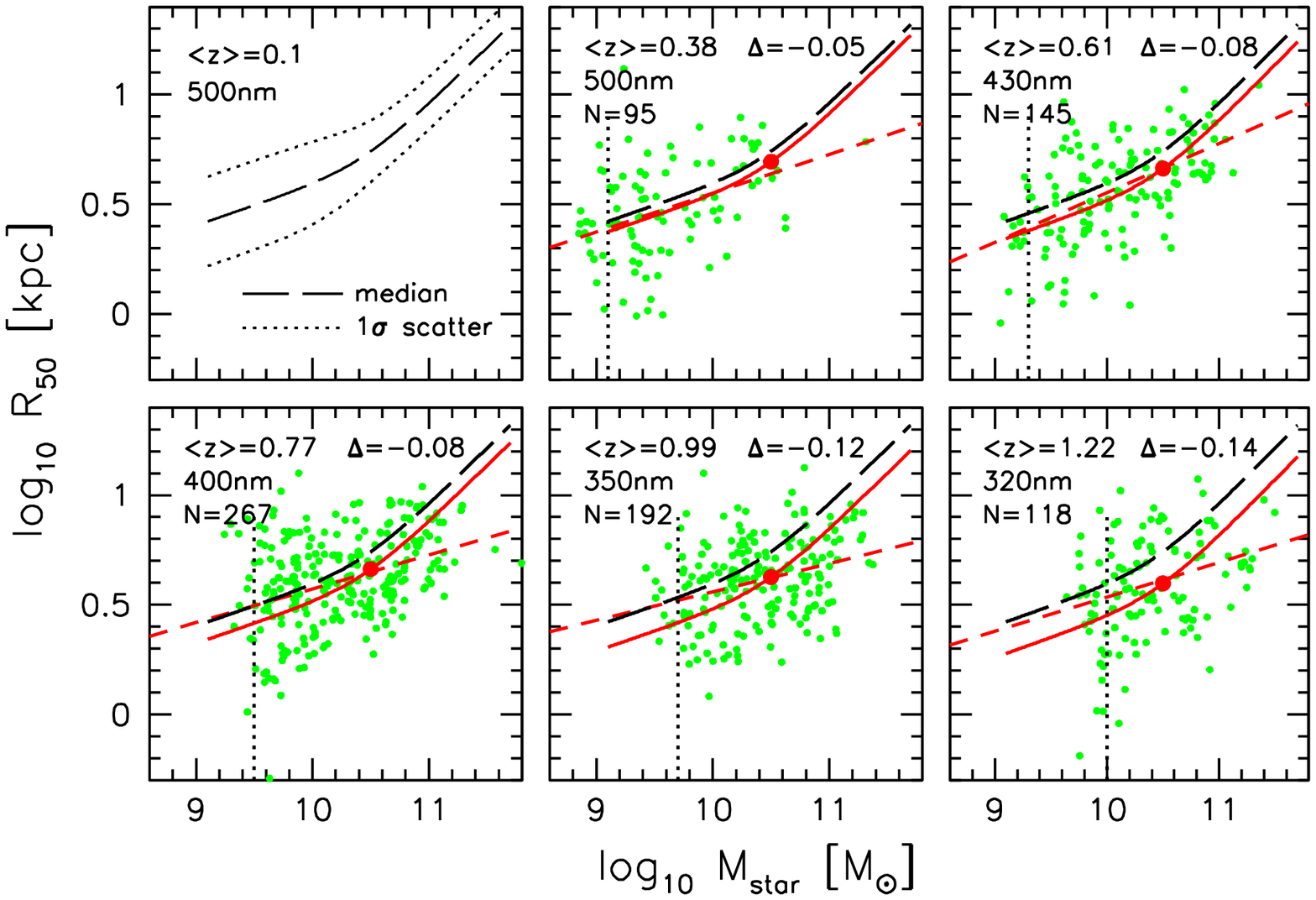,width=0.88\textwidth}}
\caption{Evolution of the disk half-light radius-stellar mass ($RM$)
  relation of disk-dominated blue-cloud galaxies from DEEP2 using
  spectroscopic redshifts.  The upper left panel shows the $RM$
  relation we derive for disk-dominated blue-cloud galaxies from
  SDSS. The long-dashed line shows the median relation, the dotted
  lines show 1$\sigma$ scatter.  In the other panels the green points
  show observations from DEEP2 using {\it HST} imaging in five
  redshift bins. The median redshift in each bin is given in the upper
  left corner of each panel, together with the central rest-frame
  wavelength of the imaging used to measure galaxy sizes, and the
  number of galaxies. The solid red line shows a fit to the DEEP2 data
  by offsetting the $z=0.1$ relation (shown by the dashed line) in
  size.  The red circle shows the median size for a stellar mass of
  $3\times 10^{10} \Msun$. The offset is given in the upper right
  corner in each panel, and is summarized in Table~\ref{tab:rmz}. The
  red short-dashed lines show a power-law fit to the data in each
  panel. The vertical dotted lines show the lower stellar mass limit
  at the upper redshift of each redshift bin.}
\label{fig:rm}
\end{figure*}

\begin{figure*}
\centerline{
\psfig{figure=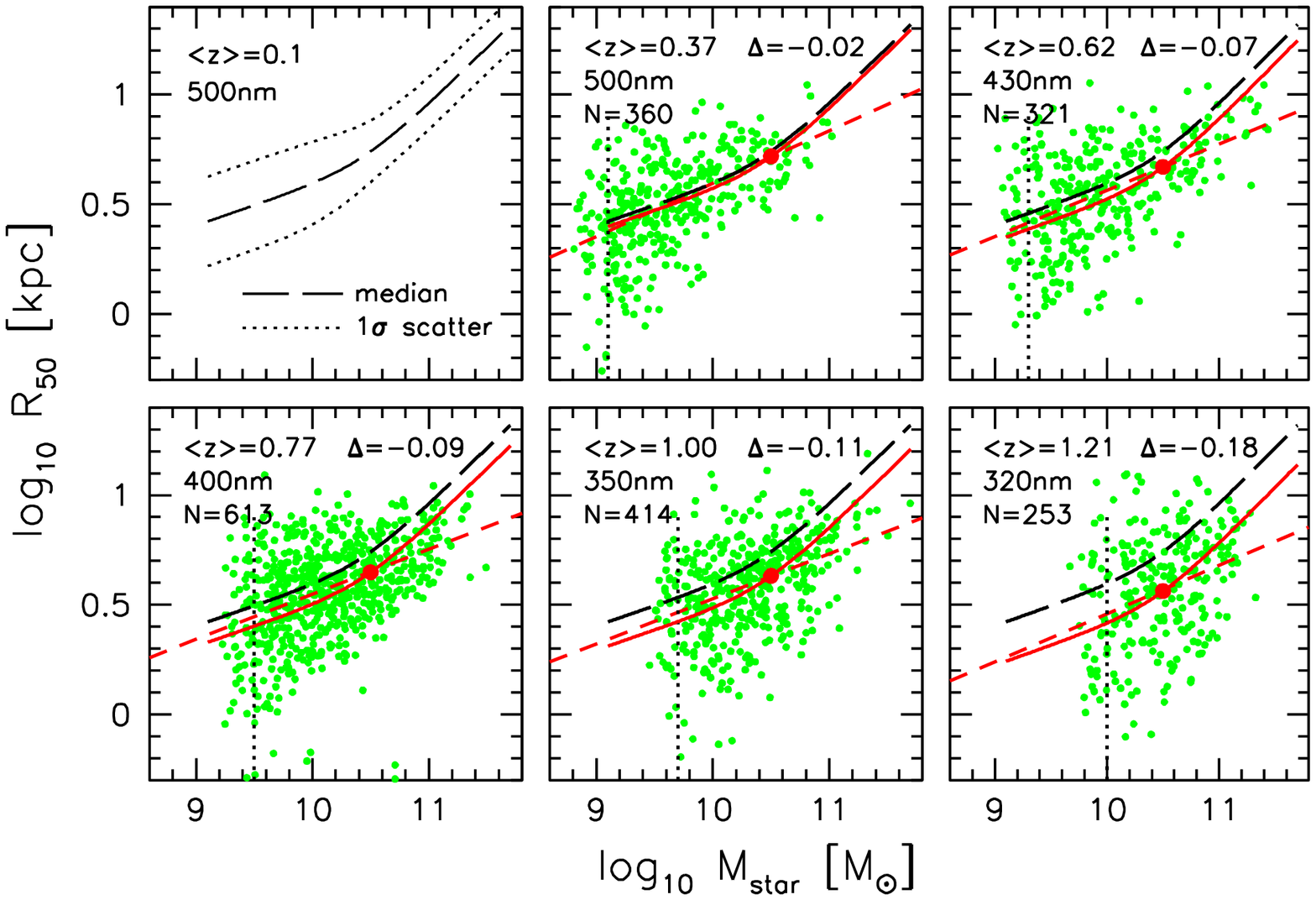,width=0.88\textwidth}}
\caption{Evolution of the disk half-light radius-stellar mass ($RM$)
  relation of disk-dominated blue-cloud galaxies from DEEP2 using
  photometric redshifts.  The upper left panel shows the $RM$ relation
  we derive for disk-dominated blue-cloud galaxies from SDSS. The
  long-dashed line shows the median relation, the dotted lines show
  1$\sigma$ scatter.  In the other panels the green points show
  observations from DEEP2 using {\it HST} imaging in five redshift
  bins. The median redshift in each bin is given in the upper left
  corner of each panel, together with the central rest-frame
  wavelength of the imaging used to measure galaxy sizes, and the
  number of galaxies. The solid red line shows a fit to the DEEP2 data
  by offsetting the $z=0.1$ relation (shown by the dashed line) in
  size.  The red circle shows the median size for a stellar mass of
  $3\times 10^{10} \Msun$. The offset is given in the upper right
  corner in each panel, and is summarized in Table~\ref{tab:rmz}. The
  red dashed lines show a power-law fit to the data in each panel. The
  vertical dotted lines show the lower stellar mass limit at the upper
  redshift of each redshift bin.}
\label{fig:rm2}
\end{figure*}

\subsubsection{SDSS comparison sample}
\label{sec:sdss}
As a low redshift comparison sample, we use data from the SDSS DR7
(Abazajian \etal 2009). We use a similar methodology to the DEEP2
sample in order to minimize systematic biases in the comparison
between high and low redshift data. Thus, we measure {\it disk
  half-light radii} with 2D bulge+disk fits using \textsc{gim2d}, performed
simultaneously on $g$ and $r$ images with the bulge and disk sizes
constrained to be the same in each filter (Simard \etal, in prep).  We
use galaxies with spectroscopic redshifts from $0.02 < z < 0.20$ and
stellar masses calculated using the relation between $r$-band
mass-to-light ratio and $(g-r)$ color from Bell \etal (2003) with an
offset of -0.1 dex to correspond to a Chabrier (2003) IMF. We select
blue-cloud galaxies for this study using the valley in the
$(g-r)$-stellar mass plane and requiring:
\begin{equation}
(g-r) < 0.71 -0.067 (\log_{10} M_{\rm star}-10).
\end{equation}
As with the DEEP2 sample we select galaxies with disk light fractions
greater than 50\% and with disk minor-to-major axis ratios greater
than 0.5 (corresponding to disk inclinations less than 60 degrees for
a zero thickness disk).  We use \textsc{gim2d} model $g$ and $r$
magnitudes, with k-corrections to $z=0$ based on SDSS petrosian
$ugriz$ magnitudes.

The SDSS spectroscopic survey is limited by apparent $r$-band
magnitude.  In the color-stellar mass plane this results in a bias
against redder galaxies at low stellar masses. In order to remove this
effect, we apply a minimum stellar mass of: $\log_{10} M_{\rm
  min}=10.4 + \log_{10} (z/0.1)$. Our final SDSS sample consists of
$\sim 130 000$ galaxies.

Fig.~\ref{fig:rmsdss} shows the disk half-light radius-total stellar
mass relation from our SDSS sample. We fit the median SDSS disk
size-mass relation with the following double power-law:
\begin{equation}
\label{eq:rm}
R = R_0 \left(\frac{M}{M_0} \right)^{\alpha} \left[\frac{1}{2} +\frac{1}{2}\left(\frac{M}{M_0}\right)^{\gamma} \right]^{(\beta-\alpha)/\gamma}.
\end{equation}
Here $\alpha$ is the slope at low masses ($M \ll M_0$), $\beta$ is the
slope at high masses ($M \gg M_0$), and $\gamma$ controls the
sharpness of the transition between the two slopes, $M_0$ is the
transition mass, and $R_0$ is the value of $R$ at $M_0$. We find
$\log_{10}(M_0/[\Msun])=10.44$, $\log_{10}(R_0/[\kpc])=0.72$,
$\alpha=0.18$, $\beta=0.52$ and $\gamma=1.8$ provide a good fit to the
data.

We assume the scatter in the SDSS disk size-mass relation is
log-normal with $s=\sigma_{\ln R|M}$, with the following relation:
\begin{equation}
\label{eq:srm}
s = s_2 + (s_1-s_2)/[1+(M/M_0)^\gamma].
\end{equation}
Here $s_1$ is the scatter at low masses, $s_2$ is the scatter at high
masses, $M_0$ is the transition mass, and $\gamma$ controls the
sharpness of the transition.  We find $s_1=0.47$, $s_2=0.27$,
$\log_{10}(M_0/[\Msun])=10.3$, and $\gamma=2.2$ provide a good fit to
the data.

For comparison, the total half-light radius-stellar mass relation for
disk galaxies (S\'ersic $n<2.5$) from the SDSS study of Shen \etal
(2003) is given by the long-dashed line, and the disk half-light size
($1.678 \times$disk scale length)-stellar mass relation from Dutton
\etal (2007), using the spiral galaxy sample of Courteau \etal (2007),
is given by the short-dashed line. The relation from Dutton \etal
(2007) is in reasonable agreement with our result considering a single
power-law was used by Dutton \etal (2007). The relation from Shen
\etal (2003) has a shallower slope at high masses, and is offset to
smaller sizes at all masses.  These differences can be understood as a
consequence of two factors: circularization and bulges. The half-light
radii used by Shen \etal (2003) are measured using circular
apertures. Since the median major to minor axis ratio of galaxy disks
is $\simeq 2$, a circular size measurement will underestimate the true
(face-on) half-light radius by a factor of $\simeq 1.4$ (0.15
dex). This effect explains most of the discrepancy at low stellar
masses. At high stellar masses there is an additional difference,
which can plausibly be explained by the increased bulge fractions in
higher mass galaxies (e.g. Dutton 2009). Since bulges tend to be
smaller than disks, total half-light radii will be smaller than disk
half-light radii.

We find that the scatter in disk sizes is mass dependent, with larger
scatter at lower masses, in qualitative agreement with Shen \etal
(2003). At low masses we find that the scatter of 0.47 in $\ln R$ is in
agreement with Shen \etal (2003), but at high masses we find a scatter
of 0.27 in $\ln R$, which is smaller than the scatter of 0.34 in $\ln
R$ reported by Shen \etal (2003). As with the difference in the slopes
at high masses, this difference is plausibly due to use of total
half-light radii by Shen \etal (2003). We note that in the simplest
\LCDM based disk formation models (e.g., Mo, Mao, \& White 1998) the
scatter in disk sizes at fixed mass is equal to the scatter in the
halo spin parameter, $\lambda$, which is $\sigma_{\ln\lambda}\simeq
0.5$ (e.g., Bullock \etal 2001b, Macci{`o} \etal 2007). This is in
rough agreement with the observed scatter at low stellar masses
($\Mstar \lta 10^{10}\Msun$), but at high stellar masses ($\Mstar \gta
10^{11} \Msun$) the observed scatter is a factor of $\simeq 2$ lower
than predicted by the simple model. This smaller than expected scatter
has been noted by previous authors (e.g., de Jong \& Lacey 2000; Shen
\etal 2003; Pizagno \etal 2005; Dutton \etal 2007). This discrepancy
may indicate that massive disk-dominated galaxies form in a subset of
haloes with a biased distribution of halo spin parameters, or that the
distribution of disk sizes has been modified by secular evolution
(e.g., Shen \etal 2003).

\subsubsection{Evolution}
The evolution of the disk half-light radius-stellar mass relation in
six redshift bins from $z=0.1$ (SDSS) up to redshift $z\simeq 1.2$
(DEEP2) is shown in Figs.~\ref{fig:rm} \&~\ref{fig:rm2}.  The upper
left panel shows fits to the SDSS disk size-stellar mass relation from
Fig.~\ref{fig:rmsdss}. For both SDSS and DEEP2 the stellar mass is the
total (i.e., disk plus bulge) stellar mass. The long-dashed line shows
the median relation, and the dotted lines show the $1\sigma$ scatter.

\begin{table}
 \centering
 \caption{Evolution of the disk half-light radius-stellar mass
   relation of blue-cloud disk-dominated galaxies relative to SDSS galaxies at
   $z=0.1$, using data from DEEP2 from Figs.~\ref{fig:rm} \&
   \ref{fig:rm2}.}
  \begin{tabular}{cccc}
\hline
\hline  
 redshift range & median $z$ & $\Delta \log_{10} R_{50} | M_{\rm star}$ & $N$ \\
\hline
\multicolumn{3}{c}{Spectroscopic Redshifts}\\
\hline
 0.2-0.5  & 0.38 & $-0.05\pm0.03 $ & 95 \\
 0.5-0.7  & 0.61 & $-0.08\pm0.02 $ & 145 \\
 0.7-0.9  & 0.77 & $-0.08\pm0.03 $ & 267\\
 0.9-1.1  & 0.99 & $-0.12\pm0.02 $ & 192 \\
 1.1-1.4  & 1.22 & $-0.14\pm0.03 $ & 118\\
\hline
\multicolumn{3}{c}{Photometric Redshifts}\\
\hline
 0.2-0.5  & 0.37 & $-0.02\pm0.01 $ & 360 \\
 0.5-0.7  & 0.62 & $-0.07\pm0.01 $ & 321 \\
 0.7-0.9  & 0.77 & $-0.09\pm0.01 $ & 613 \\
 0.9-1.1  & 1.00 & $-0.11\pm0.01 $ & 414 \\
 1.1-1.4  & 1.21 & $-0.18\pm0.02 $ & 253 \\
\hline
\hline
\label{tab:rmz}
\end{tabular}
\end{table}

Galaxies at redshifts $z>0.2$ from DEEP2 are shown as green circles.
The solid red lines show fits to the DEEP2 data obtained by
calculating the median offset in size with respect to the $z=0.1$
relation. The offsets are given in the top right corner of each panel,
and in Table ~\ref{tab:rmz}. For reference, the $z=0.1$ relation is
shown in each panel with black dashed lines. The short-dashed red
lines show power-law fits to the data in each redshift bin. These
agree well at $\Mstar\simeq 3\times 10^{10}\Msun$ (red dots in
Figs.~\ref{fig:rm} \& \ref{fig:rm2}) but suggest the size evolution
may be weaker at lower masses and stronger at higher masses.

\begin{figure}
\centerline{
\psfig{figure=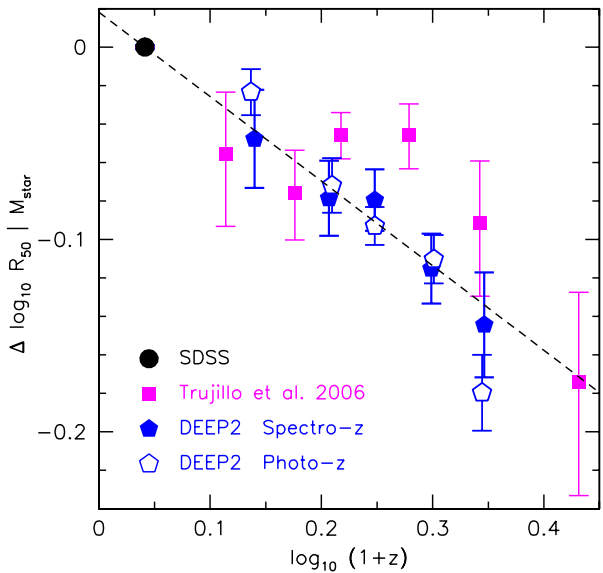,width=0.47\textwidth}}
\caption{Evolution of the disk half-light radius-stellar mass relation
  of blue/disk galaxies from redshifts $z=0.1$ to $z\simeq1.2$.  The
  pentagons show results for blue-cloud disk-dominated galaxies from
  DEEP2, where the sizes are the disk half-light sizes.  The solid
  blue pentagons show the spectro-$z$ sample (Fig.~\ref{fig:rm}), and
  the open blue pentagons show the photo-$z$ sample
  (Fig.~\ref{fig:rm2}). Both are well fit by a linear relation (dashed
  line): $\Delta \log_{10} R_{50}=0.018 -0.44 \log_{10} (1+z)$. For
  comparison, the magenta squares show data for disk-dominated
  galaxies (S\'ersic $n<2.5$) from Trujillo \etal (2006), where the
  sizes are total galaxy half-light sizes.  Error bars are $1\sigma$.}
\label{fig:rmz}
\end{figure}

The evolution of the zero point (size evolution at fixed $M_{\rm
  star}$) is summarized in Fig.~\ref{fig:rmz}. The data from DEEP2 are
shown with blue pentagons (open for photo-$z$, filled for
spectro-$z$), and the data from Trujillo \etal (2006) are shown with
magenta squares. Note that the latter are for total half-light radii,
not disk half-light radii, but nevertheless the two data sets are
consistent within the error bars. The DEEP2 data, however, shows
stronger evidence for evolution both internally and in comparison to
SDSS.  For DEEP2 the evolution of the spectro-$z$ sample is well fit
by
\begin{equation}
\Delta \log_{10} R_{50} = 0.018\pm0.002 -0.44\pm0.04 \log_{10}(1+z),
\end{equation}
where $\Delta \log_{10} R_{50}$ is the size evolution relative to $z=0.1$.

\subsection{Evolution of the size - rotation velocity relation}
\label{sec:rv}
The evolution of the size-velocity (RV) relation has been studied by
Bouch{\'e} \etal (2007) using data from the SINS survey
(F\"orster-Schreiber \etal 2009) at $z\simeq 2$ and from Courteau
(1997) at $z=0$. Using half-width half-maximum (HWHM) sizes
interpreted as exponential disk scale lengths, Bouch{\'e} \etal (2007)
found that the $RV$ relation exists at $z=2$ with the same zero-point
as that at $z=0$.

In this paper we re-determine the zero-point evolution using data from
SINS at $z\simeq 2$ adding data from Cresci \etal (2009) to those of
F\"orster-Schreiber \etal (2009). Cresci \etal (2009) determined
maximum rotation velocities for 18 galaxies at $z\simeq 2$ from the
SINS survey using kinematic modeling of 2D H$\alpha$ velocity fields.
Stellar masses were obtained from optical to NIR SED fits assuming
Solar metallicity Bruzual \& Charlot (2003) stellar population
synthesis models with a Chabrier (2003) IMF. Galaxy sizes were
measured from H$\alpha$ emission maps using two methods: Cresci \etal
(2009) give HWHM sizes, $R_{\rm HWHM}$, while F\"orster-Schreiber
\etal (2009) give circular half-light sizes, $R_{1/2}$. Both size
measurements have been made on the 18 galaxies from Cresci \etal
(2009).

\begin{figure}
\centerline{
\psfig{figure=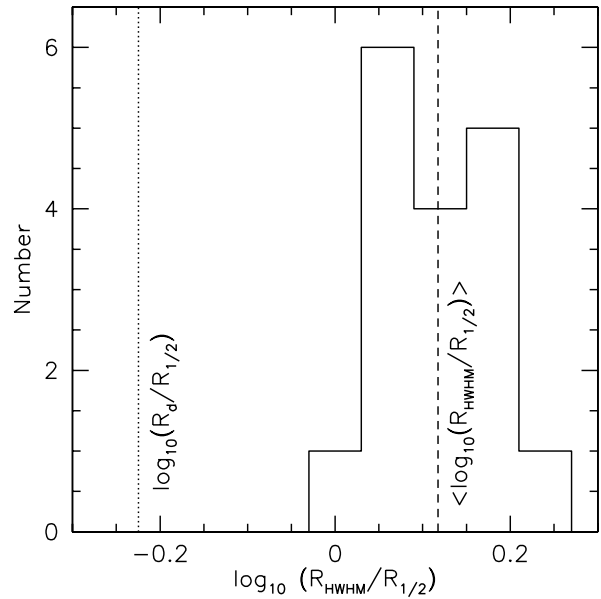,width=0.47\textwidth}}
\caption{Comparison of H$\alpha$ major-axis HWHM radii, $R_{\rm
    HWHM}$, of SINS galaxies from Cresci \etal (2009) and H$\alpha$
  half-light radii for the same 18 galaxies from F\"orster-Schreiber
  \etal (2009).  The HWHM radii are larger on average by 0.12 dex.
  Bouch{\'e} \etal (2007) claim that for an exponential distribution,
  $R_{\rm HWHM}$ equals the exponential scale length, $R_{\rm d}$, to
  within 15\%, but then $R_{\rm HWHM}$ should be 0.22 dex {\it
    smaller} than $R_{1/2}$ (see vertical line at left), contrary to
  what is observed.  Thus, either the SINS H$\alpha$ disks are
  strongly non-exponential and/or the HWHM and half-light radii
  fundamentally disagree.}
\label{fig:rr}
\end{figure}

The HWHM sizes are obtained from a linear Gaussian fit to the major
axis of the H$\alpha$ line intensity maps.  According to Bouch{\'e}
\etal (2007), the derived HWHM, once corrected for the observed
seeing, corresponds to the exponential scale length, $R_{\rm d}$, of
the disk. Using simulations of model disks these authors claim that
$R_{\rm d}$ measured this way is likely to be overestimated by no more
than 15\%.  The half-light sizes from F\"orster-Schreiber \etal (2009)
are obtained from the H$\alpha$ curves-of-growth measured in circular
apertures, and corrected for seeing.

For an exponential disk, $R_{\rm d} = 0.60 R_{1/2}$, and thus if
$R_{\rm HWHM}\equiv R_{\rm d}$ then $R_{\rm HWHM}$ should be {\it
  smaller} than $R_{1/2}$ by 0.22 dex, whereas Fig.~\ref{fig:rr} shows
that the opposite is true: the HWHM radii are 0.12 dex larger. The
fact that $R_{\rm HWHM} \simeq 1.3 R_{1/2}$ was also noted by
F\"orster-Schreiber \etal (2009).  The discrepancy may be somewhat
exaggerated, as the half-light radii are measured through circular
apertures, not along the major axis. If the galaxies studied by Cresci
\etal (2009) were a random sample of galaxies, the median disk
inclination angle would be 60 degrees, and thus the circular radii
would be too small on average by a factor of $\simeq 1.4$. However,
the distribution of disk inclinations of the galaxies in Cresci \etal
(2009) appears skewed towards low inclinations: only 3/18 galaxies
have inclination greater than 60 degrees. The median inclination is 42
degrees, which implies that the circularization effect is of order
0.07 dex.  In what follows we use both size measurements, assuming
that $R_{\rm HWHM}$ is equivalent to an exponential scale length and
$1.16\times R_{1/2}$ is equivalent to a major-axis half-light
radius. We will let consistency between various data sets determine
which, if either, size measurement is more likely to be correct.

\begin{figure*}
\centerline{
\psfig{figure=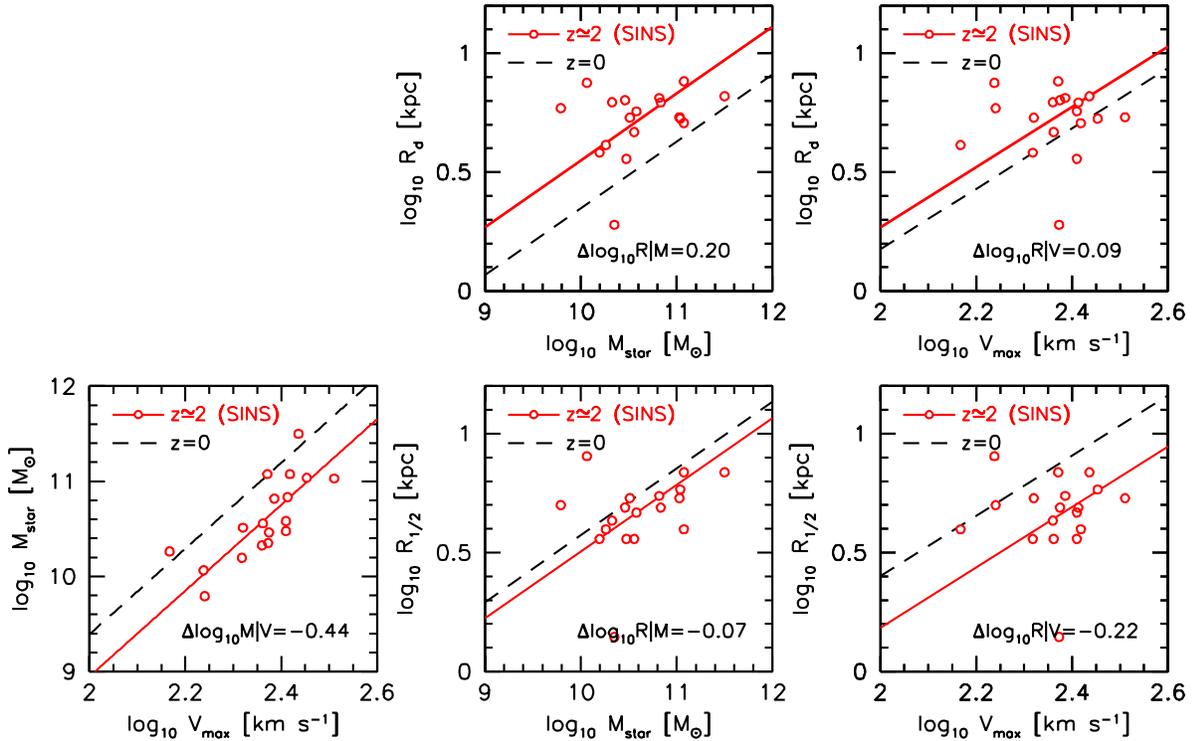,width=0.90\textwidth}}
\caption{Evolution of the size-stellar mass-rotation velocity
  relations between redshifts $z=0$ and $z\simeq 2$.  The $z=0$
  relations are from Bell \& de Jong (2001) and Dutton \etal (2007),
  while the $z\simeq 2$ data are from SINS (Cresci \etal 2009;
  F\"orster-Schreiber \etal 2009). The stellar masses and maximum
  rotation velocities are from Cresci \etal (2009). In the upper
  panels the size is the exponential disk scale length (from Cresci
  \etal 2009), whereas in the lower panels it is the half-light radius
  (from F\"orster-Schreiber \etal 2009). These two size measurements
  give different, and inconsistent, results for the evolution of
  disk-galaxy sizes.}
\label{fig:rv}
\end{figure*}

Fig.~\ref{fig:rv} shows the observed evolution of the $VMR$ relations
from $z\simeq 2$ to $z\simeq 0$. The SINS data at $z\simeq 2$ are
shown with circles. Two versions of radii are shown: HWHM
H$\alpha$ radii by Cresci \etal (2009) (upper panels) and
(de-circularized) half-light H$\alpha$ radii by F\"orster-Schreiber
\etal (2009) (lower panels).  The redshift $z=0$ relations are shown
as dashed lines and are obtained as follows.  We use the
$\Mstar-V_{\rm max}$ relation from Bell \& de Jong (2001) derived from
$K$-band luminosities using a mass dependent extinction correction,
subtracting 0.1 dex from the stellar masses to correspond to a
Chabrier (2003) IMF, and using a bisector fit:
\begin{equation}
\log_{10}\frac{\Mstar}{[10^{10}\Msun]} 
= -0.61 + 4.51\log_{10}\frac{V_{\rm max}}{[100 \kms]}.
\end{equation}
We note that the velocity used by Bell \& de Jong (2001) is not
$V_{\rm max}$, but actually $V_{\rm flat}$, the velocity in the flat
outer part of the rotation curve. However, as shown by Dutton \etal
(2010b) the TF relation derived using $V_{\rm max}$ for these galaxies
is identical to the one derived using $V_{\rm flat}$.

We use the local $I$-band disk scale length-stellar mass relation from
Dutton \etal (2007) which used an orthogonal fit:
\begin{equation}
\log_{10}\frac{R_{{\rm d},I}}{[{\rm kpc}]}
= 0.348 +0.281 \log_{10}\frac{\Mstar}{[10^{10}\Msun]},
\end{equation}
where we have adopted stellar masses from Bell \etal (2003) -0.1 dex.
Taking these $MV$ and $RM$ relations, we infer the mean relation
between $R_{{\rm d},I}$ and $V_{\rm max}$:
\begin{equation}
\log_{10}\frac{R_{{\rm d},I}}{[{\rm kpc}]}
= 0.177 +1.267\log_{10}\frac{V_{\rm max}}{[100 \kms]}.
\end{equation}
In order to compare with disk half-light radii we use
$R_{50,I}=1.678R_{{\rm d},I}$.  For simplicity we use the $I$-band
sizes to compare with the $z\simeq 2$ data. But we note that the
evolution would not change significantly, in fact it would be slightly
stronger, if we used the $V$-band $RM$ relation that we derived from
the SDSS.
 
Due to the small numbers of galaxies in the $z\simeq 2$ sample we do not
attempt to fit the slopes of the high-redshift relations. Instead we
fix the slopes to the $z=0$ values and measure the median offset of the
$z\simeq 2$ galaxies with respect to the $z=0$ relations. The values
of these offsets are given at the bottom of each panel.

The upper right panel shows the disk scale length-maximum rotation
velocity relation using the Cresci \etal (2009) SINS data. The
$z\simeq 2$ data have a marginally higher zero point ($0.09\pm 0.05$
dex) compared to the redshift zero data, which is in agreement with
the findings of Bouch{\'e} \etal (2007).  Given that the TF relation
evolves to lower stellar masses ($-0.44\pm 0.08$ dex)\footnote{Note
  that the difference in evolution of 0.05 dex between our result and
  that of Cresci \etal (2009), who use the same data, is caused by our
  different normalization of the $z=0$ relation from Bell \& de Jong
  (2001). In order to convert the stellar masses derived assuming a
  ``diet''-Salpeter IMF of Bell \& de Jong (2001) into a Chabrier
  (2003) IMF, which is used at $z\sim 2$, we adopt -0.1 dex, whereas
  Cresci \etal (2009) adopted -0.15 dex.} at fixed velocity at higher
redshifts (lower left panel), this implies that at fixed stellar mass,
sizes should be larger at higher redshifts. As is shown in the upper
middle panel, this is indeed the case, with sizes larger by
$0.20\pm0.05$ dex at $z\simeq 2$ than $z=0$. The evolution of the
$R_{\rm d}-\Mstar$ relation from the SINS Cresci \etal (2009) data is
thus of opposite sign to that of previous studies (see
\S~\ref{sec:rm}). In particular it is inconsistent with the $z\sim 2$
results from Trujillo \etal (2006) and Williams \etal (2010), who
found that disk/star forming galaxies galaxies were a factor of
$\simeq 2$ smaller at fixed stellar mass.

The lower middle and right panels show the $RM$ and $RV$ relations
using the SINS curve-of-growth H$\alpha$ half-light radii from
F\"orster-Schreiber \etal (2009).  These panels show that the
half-light sizes are {\it smaller} at a given stellar or maximum
rotation velocity, relative to $z=0$ galaxies. The offset for the $RM$
relation is $-0.07\pm0.05$ dex, while for the $RV$ relation it is
$-0.22\pm 0.06$ dex.  The evolution in the $RM$ relation is still
however, weaker than obtained by Trujillo \etal (2006), who find
$-0.28$ dex. As we discuss below in \S~\ref{sec:recon}, this remaining
difference can be accounted for by the difference between half-light
radii measured in H$\alpha$ and those measured in rest-frame $I$-band
light.


\section{Theoretical expectations}
We now compare these observational results to the predictions of
various theoretical models.
\subsection{Dark matter haloes}
The zeroth order prediction for the evolution of the $VMR$ scaling
relations of disk galaxies is given by the evolution of the $VMR$
relations of dark matter haloes.  This is obtained by assuming that:
\begin{enumerate}
\item The total mass profile of the galaxy and halo is isothermal;
\item The galaxy mass fraction, $m_{\rm gal}=M_{\rm gal}/M_{\rm vir}$,
  is a constant and independent of redshift. Here $M_{\rm gal}$ is the
  galaxy mass (i.e., the sum of stellar mass and cold gas), and
  $M_{\rm vir}$ is the total mass within the halo virial radius;
\item The galaxy spin parameter, $\lambda_{\rm gal} = (j_{\rm
    gal}/m_{\rm gal}) \lambda$, is a constant and independent of
  redshift (for a given halo). Here $j_{\rm gal}=J_{\rm gal}/J_{\rm
    vir}$ is angular momentum fraction, where $J_{\rm gal}$ is the
  total angular momentum of the galaxy and 
  $J_{\rm vir}$ is the total angular momentum within the virial radius, and 
the halo spin parameter, $\lambda$, is given by
\begin{equation}
\label{spin}
\lambda = {J_{\rm vir}  \vert E \vert^{1/2} \over  G M_{\rm
vir}^{5/2}} = \frac{\Jvir/\Mvir}{\sqrt{2}\, \Rvir \Vvir} f_c^{1/2}.
\end{equation}
Here $\Rvir$ is the virial radius, $\Vvir$ is the circular velocity at
the virial radius, $E$ is the halo's energy, and $f_{c}=1$. For more
general haloes $f_{c}$ measures the deviation of $E$ from that of a
singular isothermal sphere. For an NFW halo $f_{c}\simeq \frac{2}{3} +
(c/21.5)^{0.7}$ (see Mo, Mao, \& White 1998), where $c$ is the halo
concentration;
\item The galaxy is 100\% stars, i.e., there is no cold gas.
\end{enumerate}

\begin{figure*}
\centerline{
\psfig{figure=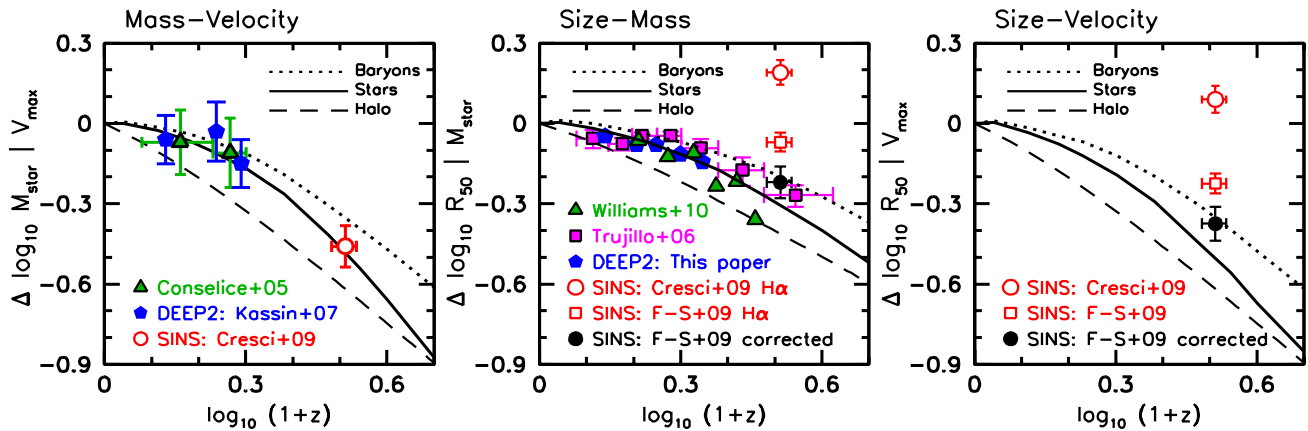,width=0.99\textwidth}}
\caption{Evolution of the circular velocity - mass -
  size relations in observations and models. The data are given
  by colored symbols, as indicated, and are described in
  \S\ref{sec:obs}.  The evolution of the stellar relations ($V_{\rm
    max}-M_{\rm star}-R_{50V}$) in a \LCDM disk galaxy evolution model
  are given by solid lines. The models and data are consistent, with
  the exception of the sizes of galaxies in the SINS survey as
  measured by Cresci \etal (2009), which appear to be a factor of
  $\simeq 3$ too large (red open circles). This discrepancy is
  partially removed when using H$\alpha$ half-light sizes from
  F\"orster-Schreiber \etal (2009) (red open squares), and fully
  removed after correcting these sizes from H$\alpha$ to rest-frame
  $I$-band (black solid circles). For comparison, the evolution of the
  baryonic ($V_{\rm max}-M_{\rm gal}-R_{\rm gal}$) and halo virial
  $(V_{\rm vir}-M_{\rm vir}-R_{\rm vir}$) relations are given by the
  dotted and dashed lines, respectively. The evolution of the
  theoretical virial, baryonic, and stellar relations are given in
  Table~\ref{tab:vmrz}.}
\label{fig:vmrz}
\end{figure*}

We refer to this model as the singular isothermal sphere (SIS) model.
Under these assumptions the disk size scales as the size and circular
velocity of the halo (e.g., Mo, Mao, \& White 1998): $R_{\rm
  d,SIS}=\lambda \Rvir/\sqrt{2} \propto \Vvir$, and the stellar mass
scales with the halo mass: $\Mstar \propto \Mvir$.

Assuming the halo is defined by spherical top hat collapse: $M_{\rm
  vir}=(4/3)\pi R_{\rm vir}^3 \Delta_{\rm vir} \rho_{\rm crit}$, then
the evolution of the $MV$, $RM$, and $RV$ relations is given by:
\begin{equation}
\frac{M_{\rm vir}(z)}{M_{\rm vir,0}} = \left[\frac{V_{\rm vir}(z)}{V_{\rm vir,0}}\right]^3 
\left[\frac{\Delta_{\rm vir}(z)}{\Delta_{\rm vir,0}}\right]^{-1/2}
\left[\frac{H(z)}{H_0}\right]^{-1}
\end{equation}

\begin{equation}
\frac{R_{\rm vir}(z)}{R_{\rm vir,0}} = \left[\frac{M_{\rm vir}(z)}{M_{\rm vir,0}}\right]^{1/3}
\left[\frac{\Delta_{\rm vir}(z)}{\Delta_{\rm vir,0}}\right]^{-1/3}
\left[\frac{H(z)}{H_0}\right]^{-2/3}
\end{equation}

\begin{equation}
  \frac{R_{\rm vir}(z)}{R_{\rm vir,0}} = \left[\frac{V_{\rm vir}(z)}{V_{\rm vir,0}}\right] 
  \left[\frac{\Delta_{\rm vir}(z)}{\Delta_{\rm vir,0}}\right]^{-1/2}
  \left[\frac{H(z)}{H_0}\right]^{-1}.
\end{equation}
The zero point evolutions are obtained by setting the leading terms on
the RHS=1.  Thus the evolution is governed by the evolution of the
Hubble parameter $H(z)$ and the halo overdensity within the virial
radius, $\Delta_{\rm vir}(z)$.  The evolution of the Hubble parameter
(in a flat cosmology) is given by
\begin{equation}
H(z) = H_0 [ \Omega_{\Lambda} + \Omega_{\rm M}(1+z)^3]^{1/2}.
\end{equation}
For the evolution of the halo overdensity, $\Delta_{\rm vir}(z)$, we
use the fitting formula from Bryan \& Norman (1998): $\Delta_{\rm vir}
= 18\pi^2 +82x -39x^2$, where $x=\Omega(z)-1$ and $\Omega(z)$ is
defined as $\Omega_{\rm M}(1+z)^3 \left[H(z)/H_0\right]^{-2}$.

Thus to first order the evolutions are determined by the evolution of
the Hubble parameter and to second order by the evolution of the halo
overdensity.  These predicted evolutions, for a \LCDM cosmology with
$\Omega_{\rm M}=0.3$ and $\Omega_{\Lambda}=0.7$, are shown by the
dashed lines in Fig.~\ref{fig:vmrz}. For the $\Mvir-\Vvir$ and
$\Rvir-\Vvir$ relations the evolution scales roughly as
$(1+z)^{-1.3}$, while for the $\Rvir-\Mvir$ relation the evolution
scales roughly as $(1+z)^{-0.8}$. The observed evolution of the
maximum circular velocity - stellar mass - half-light radius relations
is given by the colored symbols.  For the stellar mass-velocity
relation we use data from: Conselice \etal (2005, green triangles);
DEEP2 - Kassin \etal (2007, blue pentagons); and SINS - Cresci \etal
(2009, red open circle). For the size-stellar mass relation we use
data from: Trujillo \etal (2006, magenta squares); SINS - Cresci \etal
(2009, red open circle) and F\"orster-Schreiber \etal (2009, red open
square); Williams \etal (2010, green triangles); and DEEP2 -
this paper (blue pentagons). For the size-velocity relation we use
data from: SINS - Cresci \etal (2009, red open circle) and
F\"orster-Schreiber \etal (2009, red open square).  These observations
are described in more detail in \S\ref{sec:obs}.  As has been noted by
previous authors, the observed evolution of all three $VMR$ relations
is weaker than this simple prediction from dark matter haloes.

The SIS model makes a number of simplifying assumptions that are
likely to be incorrect. Firstly, total mass density profiles are not
expected to be globally isothermal. Dark matter haloes in cosmological
simulations have $V_{\rm max} > \Vvir$ (e.g., Bullock \etal 2001a).
In addition, the contribution of the baryons to the inner potential
can increase the observed $V_{\rm max}$ even further.

Using the models of Mo, Mao, \& White (1998), Somerville \etal (2008)
showed that including the expected evolution in dark halo
concentrations, as well as the contribution of baryons to the
potential, results in weaker evolution than predicted by the SIS
model. However, Somerville \etal (2008) make the assumption that disks
have no gas, and thus, that the stellar mass is equal to the baryonic
mass and the $V$-band half-light radius is equal to the baryonic
half-mass radius. In order to determine whether the stellar relations
evolve differently from the baryonic relations, one needs to follow
the evolution of stellar and gas disks in a self-consistent way. This
requires realistic cosmological simulations with gas or
semi-analytic models (SAMs). Here we use the latter.

\begin{table*}
 \centering
\begin{minipage}{0.65\textwidth}
  \caption{Evolution of the stellar, baryonic, and virial
    velocity-mass-size relations of our disk-galaxy evolution model,
    with $m_{\rm gal}=0.04$ and $\lambda=0.035$, relative to redshift
    $z=0$ as shown in Fig.~\ref{fig:vmrz}.}
  \begin{tabular}{cccccccc}
\hline
\hline  
 $z$ & $\log_{10}(1+z)$ & $\Delta \log_{10}M_{\rm star}|V_{\rm max}$ & $\Delta \log_{10} M_{\rm gal}|V_{\rm max}$ & $\Delta \log_{10} M_{\rm vir}|V_{\rm vir}$ \\ 
\hline
    0.00 &   0.000 &   0.000 &  0.000 &  0.000\\
    0.10 &   0.041 &  -0.001 &  0.006 & -0.035\\
    0.30 &   0.114 &  -0.027 & -0.014 & -0.106\\
    0.51 &   0.179 &  -0.065 & -0.039 & -0.177\\
    0.70 &   0.230 &  -0.105 & -0.066 & -0.238\\
    1.01 &   0.303 &  -0.171 & -0.115 & -0.327\\
    1.41 &   0.382 &  -0.264 & -0.190 & -0.435\\
    2.00 &   0.477 &  -0.418 & -0.307 & -0.568\\
    2.50 &   0.544 &  -0.543 & -0.394 & -0.665\\
    3.00 &   0.602 &  -0.661 & -0.472 & -0.749\\
    4.00 &   0.699 &  -0.872 & -0.610 & -0.892\\
\hline
 $z$ & $\log_{10}(1+z)$ & $\Delta\log_{10}R_{50V}|M_{\rm star}$ & $\Delta\log_{10}R_{50\rm gal}|M_{\rm gal}$ & $\Delta\log_{10}R_{\rm vir}|M_{\rm vir}$ \\
\hline
    0.00 &   0.000 &   0.000 &  0.000 & 0.000\\
    0.10 &   0.041 &   0.003 &  0.012 & -0.024\\
    0.30 &   0.114 &  -0.019 & -0.001 & -0.071\\
    0.51 &   0.179 &  -0.046 & -0.022 & -0.118\\
    0.70 &   0.230 &  -0.071 & -0.042 & -0.159\\
    1.01 &   0.303 &  -0.116 & -0.071 & -0.218\\
    1.41 &   0.382 &  -0.175 & -0.112 & -0.290\\
    2.00 &   0.477 &  -0.270 & -0.170 & -0.379\\
    2.50 &   0.544 &  -0.339 & -0.224 & -0.443\\
    3.00 &   0.602 &  -0.402 & -0.276 & -0.500\\
    4.00 &   0.699 &  -0.519 & -0.368 & -0.595\\
\hline
 $z$ & $\log_{10}(1+z)$ & $\Delta\log_{10}R_{50V}|V_{\rm max}$ & $\Delta\log_{10}R_{50\rm gal}|V_{\rm max}$ & $\Delta\log_{10}R_{\rm vir}|V_{\rm vir}$\\
\hline
    0.00 &   0.000 &   0.000 &  0.000 & 0.000\\
    0.10 &   0.041 &   0.003 &  0.010 & -0.035\\
    0.30 &   0.114 &  -0.035 & -0.016 & -0.106\\
    0.51 &   0.179 &  -0.077 & -0.046 & -0.177\\
    0.70 &   0.230 &  -0.120 & -0.069 & -0.238\\
    1.01 &   0.303 &  -0.192 & -0.121 & -0.327\\
    1.41 &   0.382 &  -0.291 & -0.197 & -0.435\\
    2.00 &   0.477 &  -0.448 & -0.306 & -0.568\\
    2.50 &   0.544 &  -0.557 & -0.395 & -0.665\\
    3.00 &   0.602 &  -0.672 & -0.479 & -0.749\\
    4.00 &   0.699 &  -0.850 & -0.614 & -0.892\\
\hline
\hline
\label{tab:vmrz}
\end{tabular}
\end{minipage}
\end{table*}

\subsection{Disk galaxy semi-analytic model}
\label{sec:sam}
In order to calculate the evolution of stellar disks as opposed to
baryonic disks, we use the disk galaxy SAM of Dutton \& van den Bosch
(2009). This model consists of disks that grow inside evolving NFW
haloes (Navarro, Frenk, \& White 1997), with structure determined from
cosmological N-body simulations (Bullock \etal 2001a; Macci{\`o},
Dutton \& van den Bosch 2008), a median spin parameter $\lambda$,
which is independent of redshift, and halo specific angular momentum
distributions from Sharma \& Steinmetz (2005), which are specified by
a parameter, $\alpha$.  The cumulative distribution of specific
angular momentum $P(<s)$, where $s$ is the specific angular momentum
in units of the total specific angular momentum, is given by
\begin{equation}
P(< s)=\gamma(\alpha,\alpha s),
\end{equation}
where $\gamma$ is the incomplete gamma function.  In this model, disks
are not formally exponential, but the stellar disks can often be well
described by an exponential profile over several scale lengths (Dutton
2009).

To build a model as close as possible to the SIS model, we maintain
assumptions (ii) and (iii) from above. As in Somerville \etal (2008),
the baryonic disk is in dynamical equilibrium inside an NFW halo which
evolves with redshift according to cosmological simulations (Bullock
\etal 2001a). The key difference is that we do not make assumption
(iv), i.e., that the disk is 100\% stars. In our model, the evolution
of the stellar and gas disks (and hence stellar and gas mass) is
governed by the radial variation of star formation, gas recycling, and
accretion.  The stellar mass is thus always less than the baryonic
mass, and the stellar disk is usually smaller in radius than the
baryonic disk. Because our model follows stellar populations as a
function of radius, we calculate the sizes of model galaxies in
optical to NIR light, as well as stellar and baryonic mass. This
allows us to test whether the sizes measured in different pass-bands
and masses are equivalent.

An additional difference between our models and those of Somerville
\etal (2008) is halo contraction. Somerville \etal (2008) assumed that
haloes contract according to the Blumenthal \etal (1986) adiabatic
contraction model. In our model, we leave the haloes uncontracted, as
models with dark halo contraction (and standard IMFs) are unable to
reproduce the zero points of the $VMR$ relations (Dutton \etal
2007). There are a variety of astrophysical processes that could
reverse the expected effect of halo contraction.  These include
dynamical friction from massive clumps of baryons (e.g., El-Zant,
Shlosman \& Hoffman 2001; Mo \& Mao 2004; Elmegreen \etal 2008;
Jardel \& Sellwood 2009), dynamical friction due to bars (e.g.,
Weinberg \& Katz 2002; Sellwood 2008), and rapid mass outflows due to
supernova feedback (e.g., Mo \& Mao 2004; Read \& Gilmore 2005;
Governato \etal 2010), so our choice of not including adiabatic
contraction is at least physically plausible as well as being
empirically motivated. As we will see from the success of our models
in matching the evolution of the $VMR$ relations, halo contraction is
also not required in order to reproduce the observed evolution of the
$VMR$ relations.

\begin{figure*}
\centerline{
\psfig{figure=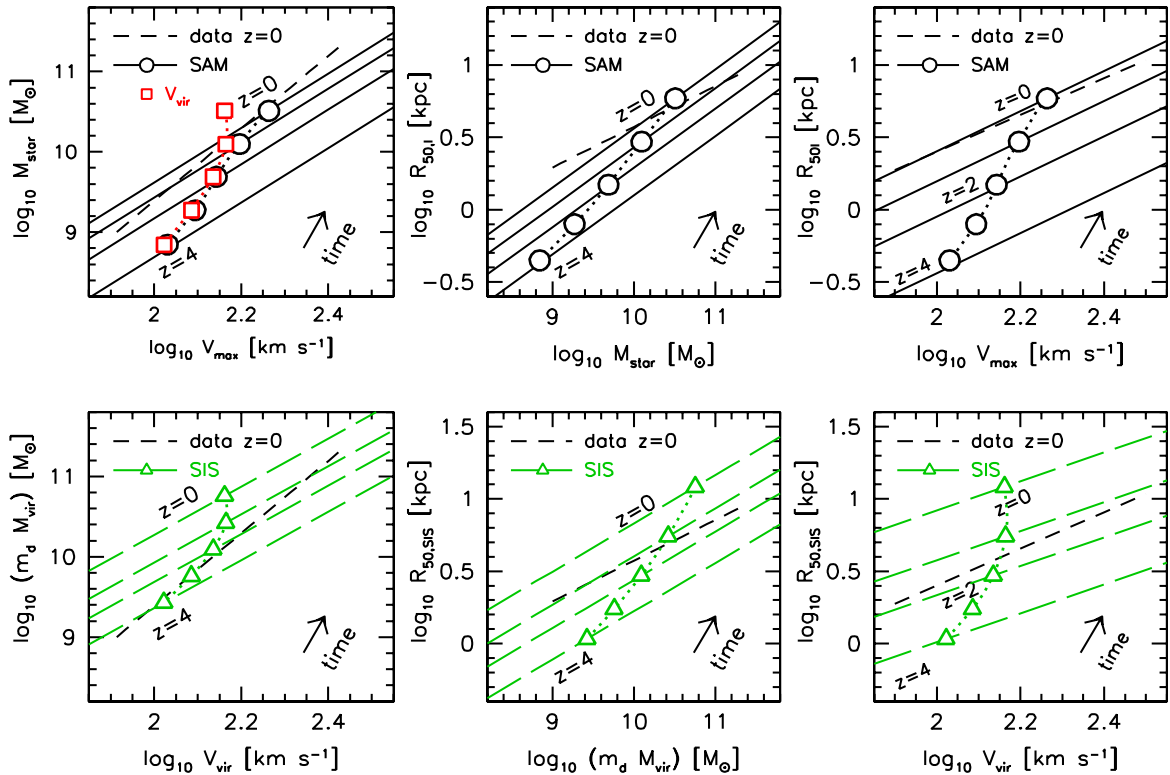,width=0.88\textwidth}}
\caption{Evolution of the maximum circular velocity - stellar mass -
  half-light radius relations in our \LCDM disk-galaxy evolution
  model.  Power-law fits to the ($V_{\rm max}-M_{\rm star}-R_{50I}$)
  relations at redshifts $z=4,2,1,0$ are given by solid black
  lines. The $z=0$ observations are given by short-dashed black lines.
  The symbols show the evolution of an individual galaxy with redshift
  $z=0$ halo virial mass of $\Mvir=10^{12}h^{-1} \Msun$ at
  $z=4,3,2,1,0$ (from lower left to top right). The black circles show
  the evolution of $V_{\rm max}-M_{\rm star}-R_{50I}$. There is
  substantial growth in these three quantities, in a roughly power-law
  from $z=4$ to $z=0$ (black dotted lines). The green triangles show
  the evolution of the SIS model, which is identical to the evolution
  of the halo virial quantities.  To facilitate comparison between the
  galaxy and virial properties, the red squares show $\Mstar-\Vvir$
  and $\Mstar-R_{\rm d,SIS}$. The weak evolution of galaxy scaling
  relations since $z\sim 1$ is due to the fact that individual
  galaxies grow roughly along the scaling relations. }
\label{fig:vmr}
\end{figure*}

In the model used here, we adopt a cosmology with $(\Omega_{\rm
  M},\Omega_{\rm \Lambda},h,\sigma_8,n)=(0.3,0.7,0.7,0.8,1.0)$, which
is close to that of the WMAP 5th year results (Dunkley \etal 2009).
We adopt a galaxy mass fraction of $m_{\rm gal}=(M_{\rm star}+M_{\rm
  gas})/M_{\rm vir} = 0.04$, a median spin parameter of
$\bar{\lambda}_{\rm gal}=\bar{\lambda}=0.035$, and a median angular
momentum shape parameter of $\bar{\alpha}=0.9$. These parameters are
motivated by observations of $m_{\rm gal}$ (Hoekstra \etal 2005;
Dutton \etal 2010b) and theoretical predictions for $\lambda$ (Bullock
\etal 2001b; Macci\`o \etal 2007) and $\alpha$ (Sharma \& Steinmetz
2005). They also result in models that roughly reproduce the observed
$VMR$ relations at $z=0$ (Fig.~\ref{fig:vmr}). However, this model
does not exactly reproduce the slopes of the local $VMR$
relations. Doing so requires either $m_{\rm gal}$ or $\lambda_{\rm
  gal}$ to vary with halo mass (e.g., Shen \etal 2003; Dutton \etal
2007).

We generate a Monte Carlo sample of 2000 galaxies with halo masses
between $\Mvir=10^{10.3}-10^{13.5} h^{-1}\Msun$, log-normal scatter in
spin parameter of $\sigma_{\ln \lambda}=0.5$ (Bullock \etal 2001b),
log-normal scatter in halo angular momentum profile of $\sigma_{\ln
  \alpha}=0.25$ (Sharma \& Steinmetz 2005), and log-normal scatter in
halo concentration of $\sigma_{\ln c}=0.25$ (Macci\`o \etal 2008).  We
determine the evolution of the zero points of the model $VMR$
relations by fitting the $MV$ and $RV$ relations for $2.1 \le
\log_{10} (V_{\rm max}/[\kms]) \le 2.5$ and the $RM$ relations for
$9.0 \le \log_{10} (M_{\rm star}/[\Msun]) \le 11.0$. We then calculate
the evolution at $\log_{10} (V_{\rm max}/[\kms]) = 2.3$ and $\log_{10}
(M_{\rm star}/[\Msun]) =10.5$.

The solid lines in Fig.~\ref{fig:vmrz} show evolution in the zero
points of the $V_{\rm max}-M_{\rm star}-R_{50V}$ (i.e., maximum
circular velocity, stellar mass, $V$-band half-light radius) relations
of this model from redshifts $z=4$ to $z=0$.  These relations show
weaker evolution than the virial relations (dashed lines), but
stronger evolution than the baryonic relations (dotted lines).  The
evolution of the theoretical virial, baryonic, and stellar relations
are given in Table~\ref{tab:vmrz}.

One interpretation of the weak evolution of the galaxy scaling
relations since $z\sim 1$ is that individual galaxies also evolve
weakly since $z\sim 1$. However, disk-galaxies were forming stars at
much higher rates in the past (e.g., Noeske \etal 2007a), and thus
stellar masses are expected to grow significantly since $z\sim
1$. Simple models for the evolution of the star formation rate-stellar
mass relation suggest that galaxies with present day stellar masses of
$3\times 10^{10}\Msun$ had stellar masses a factor of $\simeq 2.5$
lower at $z=1$ (Noeske \etal 2007b).  This amount of evolution in
stellar masses is consistent with that predicted by our
model. Fig.~\ref{fig:vmr} shows the evolution of a galaxy with present
day stellar mass of $3\times 10^{10}\Msun$. Since $z=1$ its stellar
mass has increased by a factor of $\simeq 2.5$, its half-light size
has increased by a factor of $\simeq 2$, and its maximum circular
velocity has increased by a factor of $\simeq 1.15$. In terms of
samples of galaxies, the evolution since $z=1$ is just a factor of
$\simeq 1.5$ increase in stellar masses, at fixed $V_{\rm max}$, and a
factor of $\simeq 1.3$ increase disk size, at fixed stellar mass.
Thus, the scaling relations between properties of galaxies ($V_{\rm
  max}-\Mstar-R_{50I}$) evolve only weakly since $z\sim 1$ because
individual galaxies evolve roughly {\it along} the scaling relations.
Similar theoretically based conclusions have been made previously for
the stellar mass - velocity relation (Portinari \& Sommer-Larsen 2007)
and the size-stellar mass relation (Firmani \& Avila-Reese 2009).

The differences between the evolution of the baryonic and virial $MV$
relations (dotted vs. dashed lines in Fig.~\ref{fig:vmrz}) are due
solely to evolution in the ratio between $V_{\rm max}$ and $V_{\rm
  vir}$ and not due to an evolution in the ratio between the baryonic
mass and virial mass, $m_{\rm gal}$, which is fixed to a constant in
this model. The ratio between $V_{\rm max}$ and $\Vvir$ increases
towards lower redshift due to higher concentrations in lower redshift
haloes. This results in less evolution in $V_{\rm max}$ than $\Vvir$
at fixed $M_{\rm gal}$ or $\Mvir$, and hence less evolution in $M_{\rm
  gal}$ or $\Mvir$ at fixed $V_{\rm max}$ than at fixed $\Vvir$.  The
evolution of $V_{\rm max}$ and $\Vvir$ for an individual galaxy with
redshift $z=0$ virial mass of $\Mvir=10^{12}h^{-1}\Msun$ and median
halo parameters is shown in the left panels of
Fig.~\ref{fig:vmr}. This shows that for high redshifts ($z\gta 2$)
$V_{\rm max} \simeq \Vvir$, but for low redshifts ($z\lta 1$) $\Vvir$
remains constant while $V_{\rm max}$ continues to increase. At
redshift $z=0$, $\Vmax\simeq 1.3\Vvir$, which is consistent with
recent measurements (Dutton \etal 2010b). The lower middle panel of
Fig.~\ref{fig:vmr} shows that $\Rvir$ and $\Mvir$ both continue to
increase at low redshifts. The reason $\Vvir$ remains roughly constant
while $\Rvir$ and $\Mvir$ increase is due to a trade off between the
addition of new mass (which increases $\Vvir$) and the increase of the
virial radius (which decreases $\Vvir$ because the circular velocity
of the NFW profile declines at large radius).

Since the baryonic mass fraction we adopt is only $\simeq 25\%$ of the
universal baryon fraction, it is certainly plausible that the
mechanisms responsible for making galaxy formation inefficient result
in $m_{\rm gal}$ varying with redshift. However, to first order,
variation in $m_{\rm gal}$ moves galaxies along the $VM$ relation
(e.g., Navarro \& Steinmetz 2000; Dutton \etal 2007), and thus we
expect that large changes in $m_{\rm gal}$ with redshift would be
needed in order to significantly change the evolution from that
predicted by our constant $m_{\rm gal}$ model.

The differences between the evolution of the baryonic and virial $RM$
relations in Fig.~\ref{fig:vmrz} are caused by evolution in the ratio
between baryonic and virial sizes (because the baryon mass fraction is
a constant), which are roughly given by: $R_{50 \rm gal}/\Rvir \propto
\lambda_{\rm gal} (\Vvir/V_{\rm max}) f_c^{-1/2}$. Thus as with the
$MV$ relations, the differences between the baryonic and virial $RM$
relations are driven by the evolution in $V_{\rm max}/\Vvir$. Unlike
the $VM$ relation, the baryonic $RM$ relation is sensitive to the
adopted baryon mass fraction (Dutton \etal 2007). Thus if $m_{\rm
  gal}$ decreases (or increases) with increasing redshift, this will
result in weaker (or stronger) evolution in the baryonic $RM$
relation.

The differences between the evolution of the baryonic and virial $RV$
relations are determined by evolution in both $R_{50 \rm gal}/\Rvir$
and $V_{\rm max}/\Vvir$. Evolution in the size ratio accounts of
$\simeq 2/3$ of the evolution difference, while evolution in the
velocity ratio accounts for the remaining $\simeq 1/3$. As with the
$RM$ relation, the $RV$ relation is sensitive to the galaxy mass
fraction, with a similar sign dependence.

The differences between the stellar and baryonic $MV$ relations in
Fig.~\ref{fig:vmrz} can be understood as a result of the higher cold
gas fractions at higher redshifts in our model.  We note that the
evolution in the cold gas fractions in our model is relatively modest,
with a factor of $\lta 2$ increase in gas masses at fixed stellar mass
between redshift $z=0$ and $z=2$.  As shown by Dutton, van den Bosch
\& Dekel (2010a), more general models (including cooling and outflows)
also predict weak evolution in cold gas fractions, but strong
evolution in molecular gas fractions. Both of these predictions are
consistent with recent observations (e.g., Erb \etal 2006; Daddi \etal
2010; Tacconi \etal 2010; Puech \etal 2010).

Higher cold gas fractions also contribute to different evolution in
the stellar and baryonic $RV$ and $RM$ relations. But there are also
contributions from differences between sizes in stellar mass and sizes
in optical light, which are a consequence of the inside-out nature of
stellar disk growth in our models. These differences are discussed in
more detail below.

\begin{figure*}
\centerline{
\psfig{figure=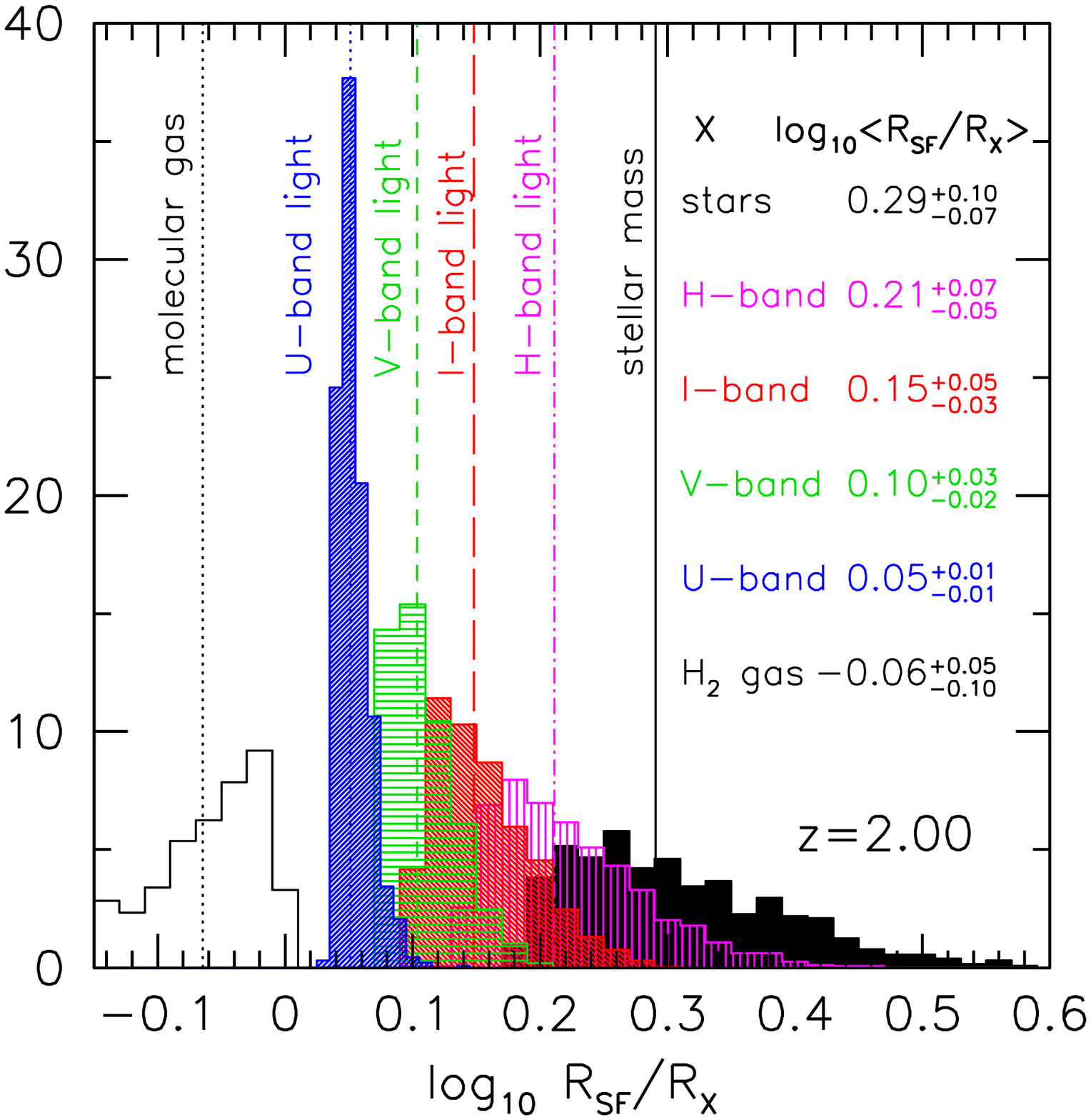,width=0.48\textwidth}
\psfig{figure=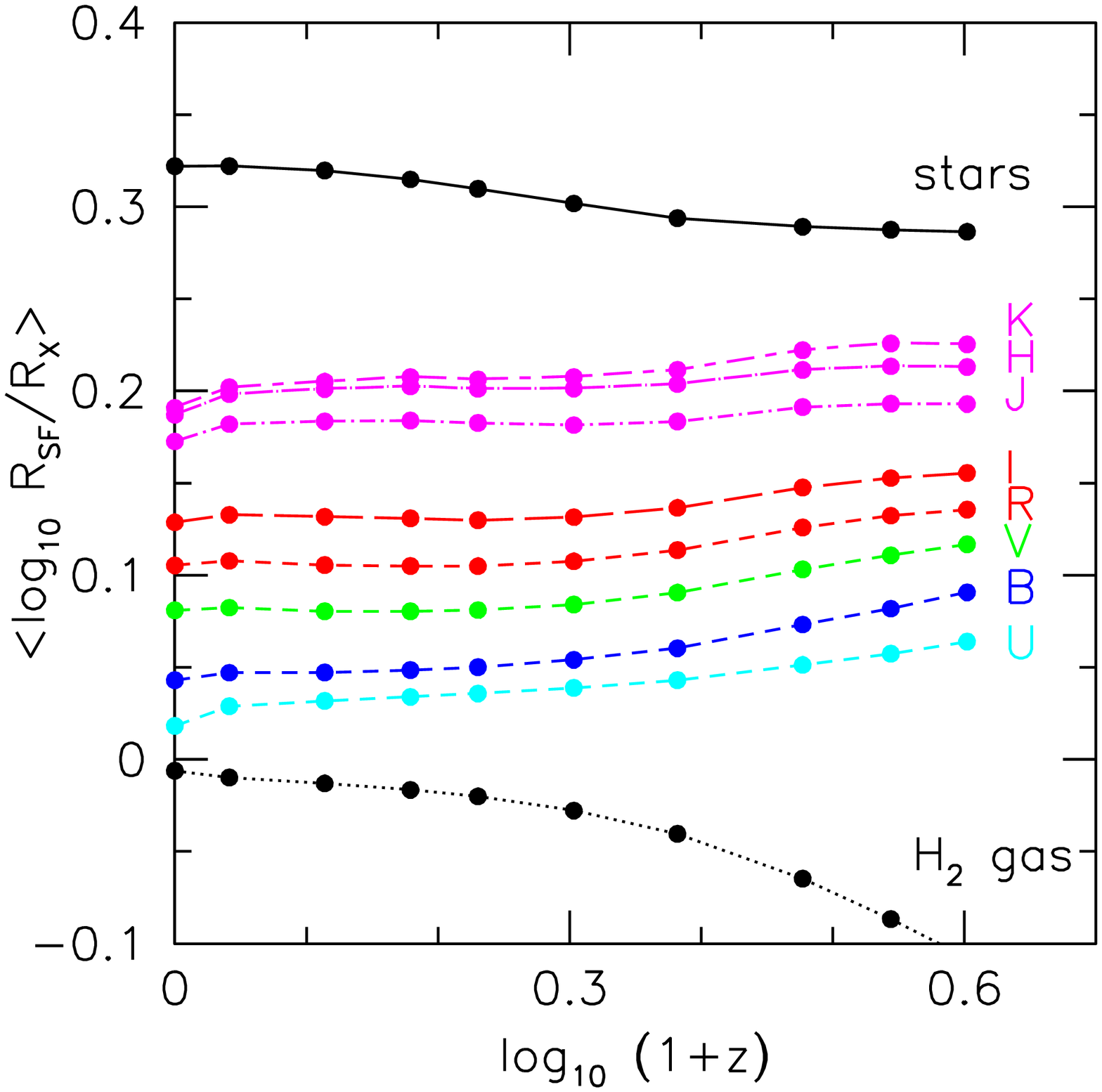,width=0.48\textwidth}
}
\caption{Ratios of galaxy sizes (half-mass or half-light) in recently
  formed stars (within the last $100 \rm Myr$), $R_{\rm SF}$, to sizes
  in various rest-frame optical-near-infrared pass bands, $R_X$, at
  redshift $z=2$ (left) and as a function of redshift (right) in a
  cosmological disk galaxy evolution model. The numbers show the
  median ratio together with the 16th and 84th percentiles of the
  distribution. The half-mass radius of the recently formed stars is,
  on average, a factor of $\simeq 2$ larger than the half-mass radius
  of the total stellar mass. This reflects the fact that stellar disks
  are growing inside out in this model.  These differences persist,
  but are not as big, when comparing $R_{\rm SF}$ to optical
  half-light radii. At $z=2$ the rest-frame $I$-band half-light sizes
  are $\simeq 0.15$ dex smaller than star formation sizes. These
  ratios evolve only weakly with redshift in our model.}
\label{fig:rrhist}
\end{figure*}

\subsection{Reconciling SINS sizes with other observations}
\label{sec:recon}
Our model nicely reproduces the observed zero point evolution of the
stellar mass Tully-Fisher ($\Mstar-\Vmax$) and size-stellar mass
relations from $z=2.2$ to $z=0$ (Fig.~\ref{fig:vmrz}).  The model also
predicts a factor of $\simeq 3$ evolution in the zero point of the
optical half-light size-velocity relation ($R_{50V}-V_{\rm max}$) over
this redshift range (solid black line in the right panel of
Fig.~\ref{fig:vmrz}). However, the $R_{\rm d}-V_{\rm max}$ data from
SINS (using disk scale lengths from Cresci \etal 2009) indicate weak
evolution in the opposite direction from $z=2.2$ to $z=0$ (red open
circle in the right panel of Fig.~\ref{fig:vmrz}). Recall that we have
derived the evolution of the $R_{\rm d}-V_{\rm max}$ relation by
comparing with the $I$-band $R_{\rm d}-V_{\rm max}$ relation at $z=0$
(see \S~\ref{sec:rv}). Since the Cresci \etal (2009) data set also
implies that higher redshift galaxies are larger at fixed stellar mass
(red open circle in the middle panel of Fig.~\ref{fig:vmrz}), this
obviously means that the evolution of the scaling relations between
different data sets are inconsistent.  The offset between the observed
$RM$ evolution from Trujillo \etal (2006) and the Cresci \etal (2009)
SINS data is also a factor of $\simeq 3$.  Thus if the Cresci \etal
(2009) disk scale lengths of the SINS galaxies could be reduced by a
factor of $\simeq 3$, then all of the data sets would be
consistent. Furthermore, the data would be consistent with our simple,
but cosmologically motivated, model for disk-galaxy evolution.

As discussed in \S\ref{sec:rv}, there are two size measurements now
available for SINS galaxies, both based on H$\alpha$ imaging: major
axis HWHM sizes by Cresci \etal (2009) (which have been interpreted as
being equivalent to disk scale lengths), and circular half-light radii
by F\"orster-Schreiber \etal (2009). In \S\ref{sec:rv} we calculated
the evolution of the SINS data at $z\simeq 2.2$ relative to the
$I$-band $RM$ and $RV$ relations at $z=0$. We found that the two size
measurements resulted in very different amounts of evolution, with the
half-light sizes giving a factor of $\simeq 2$ smaller sizes at higher
redshift. Thus using the SINS half-light sizes by F\"orster-Schreiber
\etal (2009) (red open squares in Fig.~\ref{fig:vmrz}) removes most of
the discrepancy that exists when using the Cresci \etal (2009) HWHM
sizes interpreted as disk scale lengths.

However, there is still a small discrepancy at the 0.2 dex level
between the evolution of the SINS half-light sizes with those from
Trujillo \etal (2006), Williams \etal (2010), and our models. Below
we discuss whether this difference can be explained by the difference
between sizes measured in H$\alpha$ compared to rest-frame $I$-band
light.  Since the SINS survey is not a volume limited sample of
galaxies at high-redshift, and at least some of the SINS sample was
selected on the basis of disky morphology with spatially resolved
velocity gradients, another possibility is that a selection bias
against small galaxies exists in the SINS survey.


\subsection{The relation between sizes in H$\alpha$ vs.  rest-frame optical
  light}
\label{sec:ha}
How well do galaxy sizes measured in H$\alpha$ trace those measured in
rest-frame optical stellar light?  We now address this question using
the disk SAM discussed in \S~\ref{sec:sam}. To make our comparison, we
assume that H$\alpha$ is a reliable tracer of recent star formation,
where recent star formation in our models is the star formation within
the last time step (i.e., $\simeq 30$ to $100 \rm\, Myr$). For
simplicity, we also ignore the effects of dust, which may modify the
relation between sizes measured in H$\alpha$ and optical
light. Fig.~\ref{fig:rrhist} shows the results where we compare the
half-mass radii of recent star formation, $R_{\rm SF}$, in our disk
formation model, to the half-light radii in the near-UV to near-IR
pass bands.

We find that the half-mass radius of the recent star formation is on
average a factor of $\simeq 2$ larger than the half-mass radius of the
total stellar mass.  This is a signature of the ``inside-out'' nature
of stellar disk growth in our models. The inside-out growth is due to
a combination of two factors: (1) For an individual galaxy the
baryonic disk grows with time due to an increase in the specific
angular momentum. This is cosmological inside-out growth; (2) The
density dependence of the star formation law (star formation is less
efficient at lower gas densities). This is star formation induced
inside-out growth.

Using cosmological hydrodynamical simulations of galaxies at redshift
$z \simeq 2$ Sales \etal (2009) found a similar factor of $\simeq 1.8$
difference between the half-mass sizes of the dense (star forming) gas
and the half-mass sizes of the stellar mass.  However, as we go from
stellar mass sizes to NIR sizes to optical sizes, the differences with
respect to the star formation sizes decrease. In terms of optical
half-light radii, the star formation size is just 0.1 dex higher than
$V$-band sizes and 0.15 dex higher than $I$-band sizes at $z=2$.  Thus
Fig.~\ref{fig:rrhist} suggests that the differences between H$\alpha$
and optical sizes by themselves are unlikely to explain the full
factor of 3 discrepancy between the SINS disk scale lengths and the
other observations shown in Fig.~\ref{fig:vmrz}. But, as shown by the
solid red circles, these are of the right magnitude to fully remove
the lingering small discrepancy between the F\"orster-Schreiber \etal
(2009) half-light radii and the Trujillo \etal (2006) data and our
models.

To summarize, we have shown that the radii of SINS galaxies can
plausibly be reconciled with other observational data and with
theoretical models provided proper radii are used (half-light radii,
not HWHM) and provided a small correction is applied to convert
H$\alpha$ radii to optical radii.  Ideally this agreement should be
verified directly with sizes measured in rest-frame optical light, and
preferably using the same techniques and definitions as used at lower
redshifts.  It would also be desirable to test this concordance by
measuring the evolution of the $VMR$ relations at redshifts $z\gta 1$
using larger samples of galaxies and over a wider range of masses than
current studies. The measurement of rest-frame optical sizes of
galaxies at $z\sim 2$ is currently possible with $H$-band imaging with
{\it HST}, and $H$ \& $K$-band imaging with adaptive optics from the
ground.

The right panel of Fig.~\ref{fig:rrhist} predicts that not much
evolution is expected in the size ratios between different
passbands. Thus as a test of how realistic the size ratios of our
models are at high redshift, we can compare them to observations of
size ratios at low redshift.  Fig.~\ref{fig:rrhist} shows that at
redshift $z=0$ the sizes in the $B$-band should be larger than those
in the $V$, $R$ and $H$-bands by 0.037, 0.061, and 0.145 dex,
respectively. This is consistent with observations of nearby spiral
galaxies by MacArthur, Courteau \& Holtzman (2003) who find
differences of 0.029, 0.061, and 0.127 dex. While this does not prove
that our model predicts the correct size ratios for $z=2$ galaxies, it
does provide indirect support.

\begin{table*}
 \centering
 \caption{Evolution of the TF relation in optical to near-IR
   luminosities from disk-galaxy evolution model relative to redshift
   $z=0$ as shown in Fig.~\ref{fig:tf}.}
  \begin{tabular}{cccccccc}
\hline
\hline  
 $z$ & $\log_{10}(1+z)$ & $\Delta \log_{10}L_{B}|V_{\rm max}$ & $\Delta \log_{10}L_{V}|V_{\rm max}$ & $\Delta \log_{10}L_{I}|V_{\rm max}$ & $\Delta \log_{10}L_{J}|V_{\rm max}$ & $\Delta \log_{10}L_{K}|V_{\rm max}$ \\
\hline
    0.00 & 0.000 &  0.000 & 0.000 &   0.000 &  0.000 &  0.000\\
    0.10 & 0.041 &  0.123 & 0.068 &   0.022 &  0.001 & -0.005\\
    0.30 & 0.114 &  0.236 & 0.151 &   0.076 &  0.034 &  0.021\\
    0.51 & 0.179 &  0.296 & 0.200 &   0.107 &  0.051 &  0.034\\
    0.70 & 0.230 &  0.329 & 0.225 &   0.123 &  0.060 &  0.038\\
    1.01 & 0.303 &  0.357 & 0.250 &   0.137 &  0.060 &  0.036\\
    1.41 & 0.382 &  0.374 & 0.263 &   0.139 &  0.048 &  0.018\\
    2.00 & 0.477 &  0.353 & 0.234 &   0.094 & -0.010 & -0.051\\
    2.50 & 0.544 &  0.319 & 0.191 &   0.043 & -0.070 & -0.118\\
    3.00 & 0.602 &  0.279 & 0.143 &  -0.016 & -0.132 & -0.186\\
    4.00 & 0.699 &  0.192 & 0.047 &  -0.123 & -0.245 & -0.311\\
\hline
\hline
\label{tab:tf}
\end{tabular}
\end{table*}

Another prediction of the model is that sizes in molecular gas should
be only slightly larger than sizes in H$\alpha$ or UV light. This is a
consequence of the almost linear relation between star formation rate
surface density and molecular gas surface density in our model (See
Fig. A1, Dutton \etal 2010a). At higher molecular
gas densities the slope of the star formation law in our model is
steeper than unity, with an asymptotic value of 1.4, in agreement with
the standard Kennicutt-Schmidt relation (Kennicutt 1998). Since
molecular gas disks are predicted to be higher density at higher
redshifts, this results in a slight evolution in the ratio between
molecular gas sizes and star formation rate sizes in our model (right
panel, Fig.~\ref{fig:rrhist}). The similarity between molecular gas
sizes and UV sizes has already been observed for a handful of star
forming galaxies at redshift $z\sim 1.5$ (Daddi \etal 2010), which
provides further support for our model.

\subsection{The evolution of the Tully-Fisher relation in optical to
  near-IR luminosities} 

\begin{figure}
\centerline{
\psfig{figure=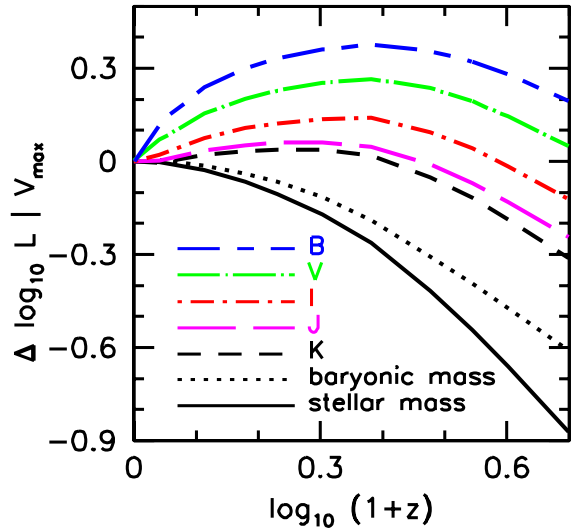,width=0.46\textwidth}}
\caption{Evolution of the Tully-Fisher relation using luminosities
  measured in rest-frame optical to near-IR pass bands in our
  disk-galaxy semi-analytic model. At fixed circular velocity,
  luminosities increase out to redshift $z\simeq 1.5$, with stronger
  evolution in bluer passbands than the near-IR. For comparison, the
  evolution of the stellar and baryonic Tully-Fisher relations are
  shown with solid and dotted lines, respectively, which are the same
  as those in the left panel of Fig.~\ref{fig:vmrz}. }
\label{fig:tf}
\end{figure}

The evolution of our model Tully-Fisher relations in optical to
near-IR luminosities is shown in Fig.~\ref{fig:tf} and in
Table~\ref{tab:tf}. Contrary to the stellar and baryonic TF relations,
which show a {\it decrease} in mass at fixed circular velocity (solid
and dotted lines), the luminosity TF relations show an {\it increase}
in luminosity at fixed circular velocity between today and redshift
$z\sim 1$. Luminosities in bluer pass-bands show stronger evolution
than redder pass-bands. In the $B$-band, the evolution is 0.36 dex (or
0.9 magnitudes) since $z=1$, while in the $K$-band, the evolution is
just 0.04 dex (or 0.1 magnitudes) since $z=1$. A similar theoretical
prediction of $0.85$ mag for the evolution of the $B$-band TF relation
since $z=1$ has been shown by Portinari \& Sommer-Larsen (2007) using
cosmological hydrodynamical simulations. Using a semi-analytic model
Firmani \& Avila-Reese (2003) find differences between the evolution
of the $B$-band and $H$-band TF relations which are qualitatively
similar to what we find in our models, though in detail there are
differences in the absolute evolution.

Most observations find no or weak evolution in the $J$- and $K$-band
TF relations since $z\sim 1$ (Conselice \etal 2005; Weiner \etal
2006b; Flores \etal 2006; Fern{\'a}ndez Lorenzo \etal 2010), which is
in good agreement with our model (but see Puech \etal (2008) who find
0.6 magnitudes of brightening between $z\sim 0.6$ and $z=0$). For the
$B$-band TF relation, a wide range of evolution has been reported, but
almost all studies find a brightening in $B$-band luminosities at
higher redshifts (e.g., Vogt \etal 1996,1997; Simard \& Pritchet 1998;
Ziegler \etal 2002; B\"ohm \etal 2004; Bamford \etal 2005,2006; Weiner
\etal 2006b; Chiu \etal 2007; Fern{\'a}ndez Lorenzo \etal 2010). Most
studies are consistent with a brightening of $1\pm 0.5$ magnitudes
from $z=0$ to $z\sim 1$, and thus in agreement with our model.

Finally, we note that no evolution in the $J$- or $K$-band TF relation
does not imply no evolution in the stellar mass TF relation. In our
model the stellar masses at fixed velocity decrease towards higher
redshifts, but the light-to-mass ratios increase monotonically towards
higher redshifts (because the stellar populations are progressively
younger). The net effect is very weak evolution in $J$- and $K$-band
TF relations between $z\sim 2$ and $z=0$.


\section{Summary}
\label{sec:conclusion}

We study the evolution in the zero points of the relations between
maximum rotation velocity, $V$, stellar mass, $M$, and rest-frame
optical disk size, $R$, of disk-galaxies in the context of \LCDM based
galaxy formation models.

Using data from the DEEP2 survey to measure the evolution of the $RM$
relation since $z=1.2$, together with published results from Conselice
\etal (2005), Trujillo \etal (2006), Kassin \etal (2007), and the SINS
survey (Cresci \etal 2009; F\"orster-Schreiber \etal 2009), we show
that there is a consistent observational picture, with one exception,
for the evolution of the $VMR$ relations from redshifts $z\simeq 2.2$
to $z\simeq 0$.  The exception is that the H$\alpha$ exponential disk
scale lengths of galaxies from the SINS survey measured by Cresci
\etal (2009), appear to be a factor of $\simeq 3$ higher at fixed $M$
than found by other observations. This apparent discrepancy can be
traced to two factors.

Firstly, Bouch{\'e} \etal (2007) and Cresci \etal (2009) measure HWHM
(half-width half-max) sizes and interpret these as exponential disk
scale lengths. However, the H$\alpha$ half-light radii from
F\"orster-Schreiber \etal (2009) of the same galaxies do not support
this interpretation. Instead, they suggest that the HWHM overestimates
the disk scale length by a factor of $\simeq 2$.

Secondly, using a \LCDM based disk galaxy formation model (Dutton \&
van den Bosch 2009) we show that disk-galaxies at redshifts $z=0-3$
are expected to have half-mass radii of recently formed stars, $R_{\rm
  SF}$, a factor of $\simeq 2$ higher than the half-mass radius of the
total stellar mass, $R_{\rm star}$. This is a direct consequence of
inside out disk growth. In terms of optical half-light radii, our
models predict that, $R_{\rm SF}\simeq 1.4 R_I \simeq 1.25 R_V$ with
little dependence on redshift.

Additionally, since the SINS survey is not a volume limited sample of
galaxies at high-redshift, and at least some of the SINS sample was
selected on the basis of disky morphology with spatially resolved
velocity gradients, another possibility (which we do not invoke here)
is that a selection bias against small galaxies exists in the SINS
survey. In order to rule-out this possibility the size-mass relation
of SINS galaxies needs to be compared to that of a volume limited
control sample, with the same methods used to derive sizes and masses.

We further show that the observed evolution of the $VMR$ relations is
consistent with a simple $\Lambda$CDM-based model of disks growing
inside evolving NFW dark-matter haloes.  This model adopts a constant
disk-to-halo mass fraction of $m_{\rm d}=0.04$, and a median spin
parameter of $\lambda=0.035$, independent of redshift.  The galaxy
mass fraction is consistent with observations at low redshifts
(Hoekstra \etal 2005; Dutton \etal 2010b), and the spin parameter is
consistent with expectations from cosmological simulations (e.g.,
Bullock \etal 2001b; Macci{\`o} \etal 2007).  While our model is
certainly an over-simplification of disk-galaxy evolution, it
demonstrates that there is no need to invoke abnormally high spin
parameters to explain the scaling relations of star-forming
disk-galaxies at high redshifts ($z\simeq 2$) as has been claimed by
Bouch{\'e} \etal (2007) and Burkert \etal (2009).

The weak evolution of the galaxy scaling relations since $z\sim 1$ in
our model is due to the fact that individual galaxies grow roughly
along the scaling relations, and not due to weak evolution in the
properties of individual galaxies themselves. Similar conclusions for
the stellar mass-velocity relation have been reached by Portinari \&
Sommer-Larsen (2007) using cosmological simulations, and for the
size-stellar mass relation by Firmani \& Avila-Reese (2009) using a
semi-analytic model similar to the one used here.  For example, for a
galaxy with present day stellar mass of $3\times 10^{10}\Msun$, since
$z=1$, its stellar mass has increased by a factor of $\simeq 2.5$, its
half-light size has increased by a factor of $\simeq 2.0$, and its
maximum circular velocity has increased by a factor of $\simeq
1.15$. In terms of samples of galaxies, the evolution is just a factor
of $\simeq 1.5$ increase in stellar mass, at fixed $V_{\rm max}$, and
a factor of $\simeq 1.3$ increase in disk size at fixed stellar mass,
since $z=1$. Evolution in stellar mass by a factor of $\simeq 2.5$ for
galaxies with present day stellar masses of $3\times 10^{10}\Msun$ is
also consistent with the evolution of the star formation rate -
stellar mass relation since $z=1$ (Noeske \etal 2007b).

In our models the evolution of the stellar scaling relations is {\it
  stronger} than that of the baryonic scaling relations (maximum
circular velocity, baryonic mass, baryonic half-mass size).  This is
due to a combination of the inside-out nature of stellar disk growth
in \LCDM cosmologies, coupled to a decrease in cold gas fractions with
cosmic time.

In our models the evolution of the baryonic scaling relations is {\it
  weaker} than that of the virial scaling relations of dark matter
haloes, assuming a constant galaxy mass fraction. For example, the
baryonic TF and baryonic size-velocity relations evolve by just
$\simeq 0.11$ dex from $z=1$ to $z=0$, whereas the corresponding
relations between halo virial quantities evolve by $\simeq 0.33$ dex
from $z=1$ to $z=0$.  This difference can be understood as a
consequence of the ratio between maximum circular velocity and virial
circular velocity, $V_{\rm max}/\Vvir$, increasing towards lower
redshifts. This in turn is largely a consequence of the increase in
halo concentrations with time (e.g., Bullock \etal 2001a).

While we have shown that there is a consistent observational and
theoretical picture for the evolution of the $VMR$ relations out to
redshift $z\sim 2$, there is much room for progress.  Our theoretical
model makes a number of simplifying assumptions which are unlikely to
be correct in detail.  We assume that the galaxy mass fraction and
galaxy spin parameters do not evolve with time. While this is a
reasonable assumption to start with, we note that this is not a
natural outcome of our model when we include cooling and outflows. Our
model also assumes that galaxy disks are smooth, and 100\% supported
by rotation. This assumption is valid in the local universe, and
perhaps up to redshift $z\sim 1$, but may break down at higher ($z\gta
2$) redshifts, especially for lower mass galaxies.

Current observational samples at $z \gta 1$ are small and/or subject
to measurement uncertainties and selection biases. For the $RM$
relation, it will be possible, in the near future, to measure robust
rest-frame optical sizes at $z\lta 2.5$ using large NIR surveys with
{\it HST}/Wide Field Camera 3. Measuring maximum circular velocities is currently a
challenge at high redshifts, as it is limited to ground based NIR
spectroscopy. These observations are typically seeing limited, which
complicates the measurement of maximum rotation velocities. However,
coming generations of adaptive optics-assisted ground based telescopes
will open up the field in concert with the Atacama Large
Millimeter/submillimeter Array (ALMA) and the Square Kilometer Array
(SKA). With this suite of frontier instrumentation, it will be
possible to measure resolved rotation curves and gas density profiles
in molecular and atomic gas out to high redshifts.  Coupled to
observations of stellar masses and sizes, these observations will
enable the evolution of the baryonic $VMR$ scaling relations over a
large fraction of cosmic time to be measured. These scaling relations
will provide a complete set of observational constraints with which to
test models of disk-galaxy formation.

\section*{Acknowledgements} We thank Nicolas Bouch{\'e} for useful
discussions regarding the SINS data.  A.A.D.  acknowledges financial
support from a CITA National Fellowship, from the National Science
Foundation grant AST-08-08133, and from Hubble Space Telescope grants
GO-10532.02-A and GO-11206.02-A.

The DEEP2 survey was initiated under the auspices of the NSF Center
for Particle Astrophysics. Major grant support was provided by
National Science Foundation grants AST 95-29098, 00-711098, 05-07483,
and 08-08133 to UCSC and AST 00-71048, 05-07428, and 08-07630 to UCB.
The DEEP2 survey has been made possible through the dedicated efforts
of the DEIMOS instrument team at UC Santa Cruz and support of the
staff at Keck Observatory.  The {\it HST} ACS mosaic in EGS was
constructed by Anton Koekemoer and Jennifer Lotz and was funded by
grant HST-AR-01947 from NASA.  Finally, we recognize and acknowledge
the highly significant cultural role and reverence that the summit of
Mauna Kea has always had within the indigenous Hawaiian community; it
has been a privilege to be given the opportunity to conduct
observations from this mountain.


\label{lastpage}

\end{document}